\documentclass[twocolumn]{aastex62}

\newcommand{\rhocirc}{\rho_{\rm circ}}
 \newcommand{\prob}{{\rm prob}}

\newcommand{\TESS}{\emph{TESS}\xspace}

\newcommand{\kep}{\emph{Kepler}\xspace}
\usepackage{xspace}
\usepackage{amsmath}

\shorttitle{Impact Parameter}
\shortauthors{Dawson}

\begin{document}

\title{Robustly detecting changes in warm Jupiters' transit impact parameters}

\author{Rebekah I. Dawson}
\affiliation{Department of Astronomy \& Astrophysics, The Pennsylvania State University; Center for Exoplanets and Habitable Worlds, The Pennsylvania State University; {\tt rdawson@psu.edu}.}

\begin{abstract}
Torques from a mutually inclined perturber can change a transiting planet's impact parameter, resulting in variations in the transit shape and duration. Detection of and upper limits on changes in impact parameter yield valuable constraints on a planetary system's three dimensional architecture. Constraints for warm Jupiters are particularly interesting because they allow us to test origins theories that invoke a mutually inclined perturber. Because of warm Jupiters' high signal-to-noise transits, changes in impact parameter are feasible to detect. However, here we show that allowing the impact parameter to vary uniformly and independently from transit to transit leads to incorrect inferences about the change, propagating to incorrect inferences about the perturber. We demonstrate that an appropriate prior on the change in impact parameter mitigates this problem. We apply our approach to eight systems from the literature and find evidence for changes in impact parameter for warm Jupiter Kepler-46b. We conclude with our recommendations for light curve fitting, { including when to fit impact parameters vs. transit durations}.
\end{abstract}

\section{Introduction}

\begin{figure}[b!]
\includegraphics[width=\columnwidth]{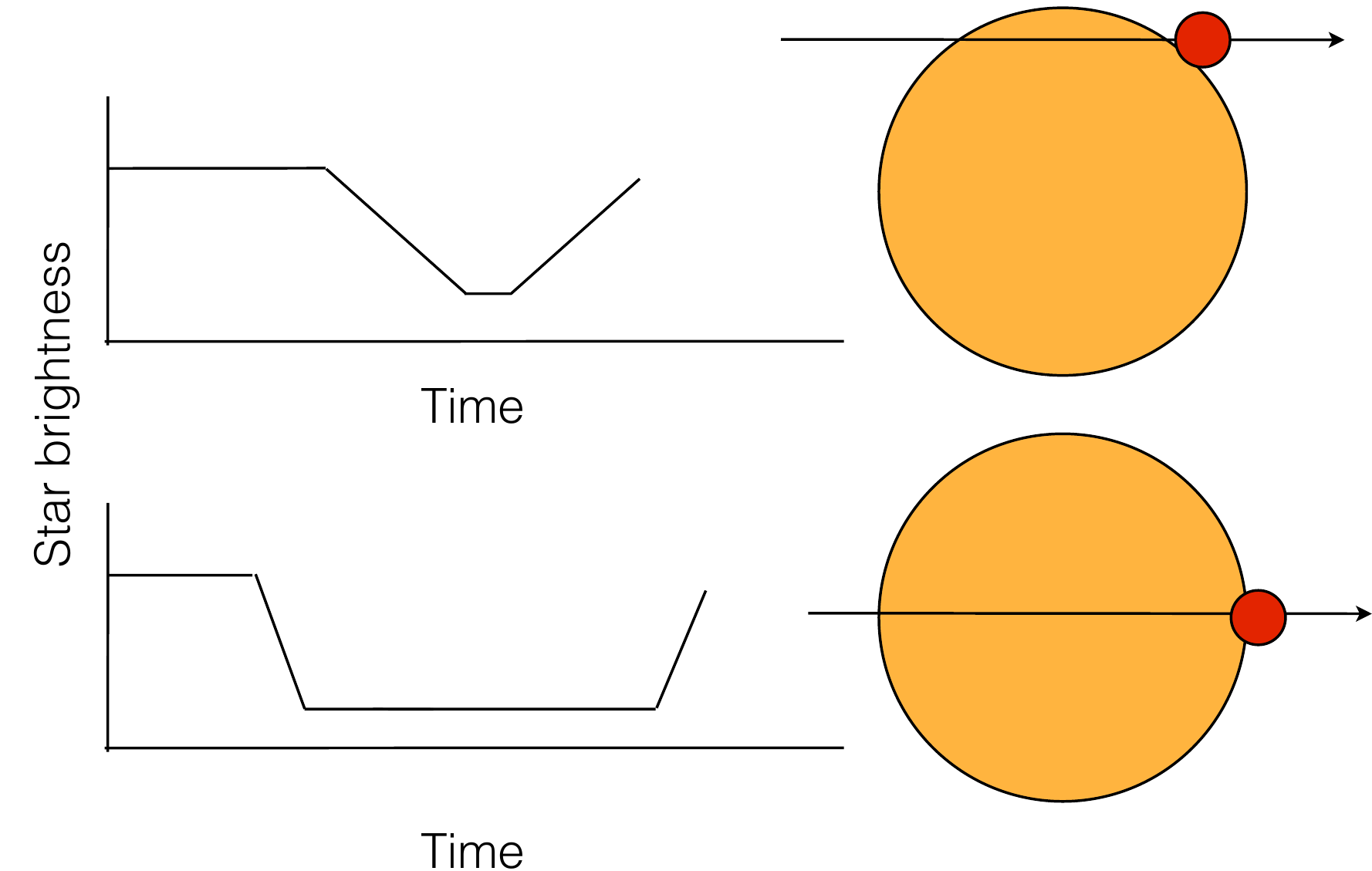}
\caption{ Schematic: the transit impact parameter affects the shape and duration of the transit. Top: a large impact parameter corresponds to a short chord and shorter transit duration. Bottom: a small impact parameter ($b=0$) corresponds to a maximal chord length and transit duration. \label{fig:cart}
}
\end{figure}

\begin{figure}[b!]
\includegraphics[width=\columnwidth]{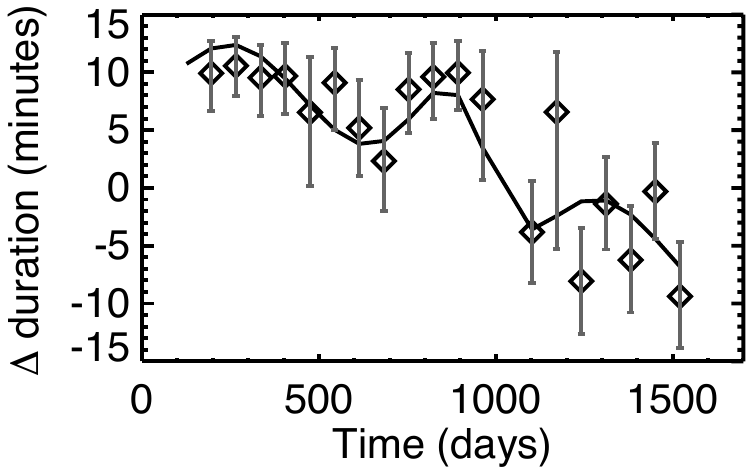}
\caption{Injected and recovered change in transit duration variations of a warm Jupiter (mass$ = 2 M_{\rm Jup}$, $P$=70 day, $e=0.47$) torqued by a mutually inclined outer planet (mass $7 M_{\rm Jup}$, $a = 2$ AU, $e=0.1$, $i_{\rm mut}=60^\circ$). Transits were injected (solid line) into the out-of-transit long cadence data of a 14.8 Kepler magnitude warm Jupiter host and recovered/fit using our pipeline (diamonds with error bars). \label{fig:ex}
}
\end{figure}

When a transiting planet is torqued by a body on a mutually inclined orbit, its transit shape and duration change (Figure \ref{fig:cart}). These changes give us a rare handle on the three-dimensional architectures of planetary systems, which are essential for testing theories of their dynamical origin. Such constraints are especially meaningful and achievable for a class of planets known as warm Jupiters, giant planets with $10--200$ day orbital periods. Popular theories for the origins of warm Jupiters -- particularly those on elliptical orbits -- predict they will be accompanied by a mutually inclined giant planet at $\sim1-5$ AU (e.g., \citealt{dong14,daws14b,petr16,ande17}). The required orbital properties for these outer planets result in changes in shape and duration that are feasible to detect with \kep light curves for warm Jupiters. For example, a warm Jupiter on a 70 day orbit torqued by a seven Jupiter mass, $60^\circ$ mutually inclination companion at 2 AU would exhibit a 30 minute change in its transit duration over the duration { of the prime} \kep Mission. When we inject such transits into KOI-3309, a warm Jupiter host with a typical Kp magnitude of 14.8, we recover the duration of each individual transit with 3-5 minute precision (Fig \ref{fig:ex}). Several studies of warm Jupiters have considered measurements or upper limits on changes in impact parameter and/or transit duration in studying the system's three dimensional architecture (e.g., \citealt{nesv12,nesv13,nesv14,daws14,masu17,mill17}).

Changes in transit shape and duration can result from a change in either the impact parameter (the distance of the transit chord from the center of the star; Fig. \ref{fig:cart}) or the transit speed, but we expect the change in impact parameter to dominate. The well-separated perturbers invoked as warm Jupiters' putative companions cause secular variations in the warm Jupiter's sky-plane inclination and eccentricity on timescales of thousands of years or longer.  Consider a transiting planet located at 0.5 AU from its sun-like star with a sky-plane inclination of { 89.725}$^\circ$. A mere 0.1$^\circ$ (0.0017 rad) tweak in the sky-plane inclination changes the impact parameter from 0.52 to 0.70, resulting in hefty { 17\%} change in the transit duration. To get an equivalent change in the duration caused by the transit speed would require a full $180^\circ$ precession for eccentricity $e=0.1$, a $35^\circ$ precession for $e=0.5$, or an increase in eccentricity from 0.1 to 0.25 (or 0.5 to 0.6).  Therefore, using the prior knowledge that the change in impact parameter dominates, we can obtain the most precise constraints on the perturbing companion by allowing for a change in impact parameter while keeping the transit speed constant. 

However, here we will show that fitting one transit speed (or, equivalently, planet-star separation or light curve stellar density) for all transits while allowing each transit to have its own impact parameter leads to flawed inferences about transit parameters. The inferred values can differ from the truth at the tens of sigma level. These incorrect parameters translate into incorrect constraints on the perturbing companion. In Section \ref{sec:orig}, we demonstrate this problem and explain its origin. In Section \ref{sec:miti}, we show that an appropriate prior on the change in impact parameter mitigates the problem. Conversely, a uniform prior corresponds to unphysical assumptions about the gravitational dynamics. { We also discuss when to fit impact parameters vs. transit durations.} In Sections \ref{sec:appl1} and \ref{sec:appl2}, we apply our approach to \kep and TESS systems from the literature and compare with previous analyses { (most of which were not subject to the bias described here)}. We summarize our findings, including recommendations for light curve fitting, in Section \ref{sec:disc}.

\section{Origin of flawed inferences from transit duration variations}
\label{sec:orig}

Here we show that when we fit a planet's transit light curve and assume a uniform prior on the magnitude of the variation in impact parameter from transit to transit, we make incorrect inferences about transit parameters. These incorrect parameters lead to incorrect inferences about the presence and properties of a perturbing body. In this section, we explain the origin of the flawed inferences from transit duration variations. 

\subsection{Overview of light curve inference}

\begin{figure}[htbp]
\includegraphics[width=\columnwidth]{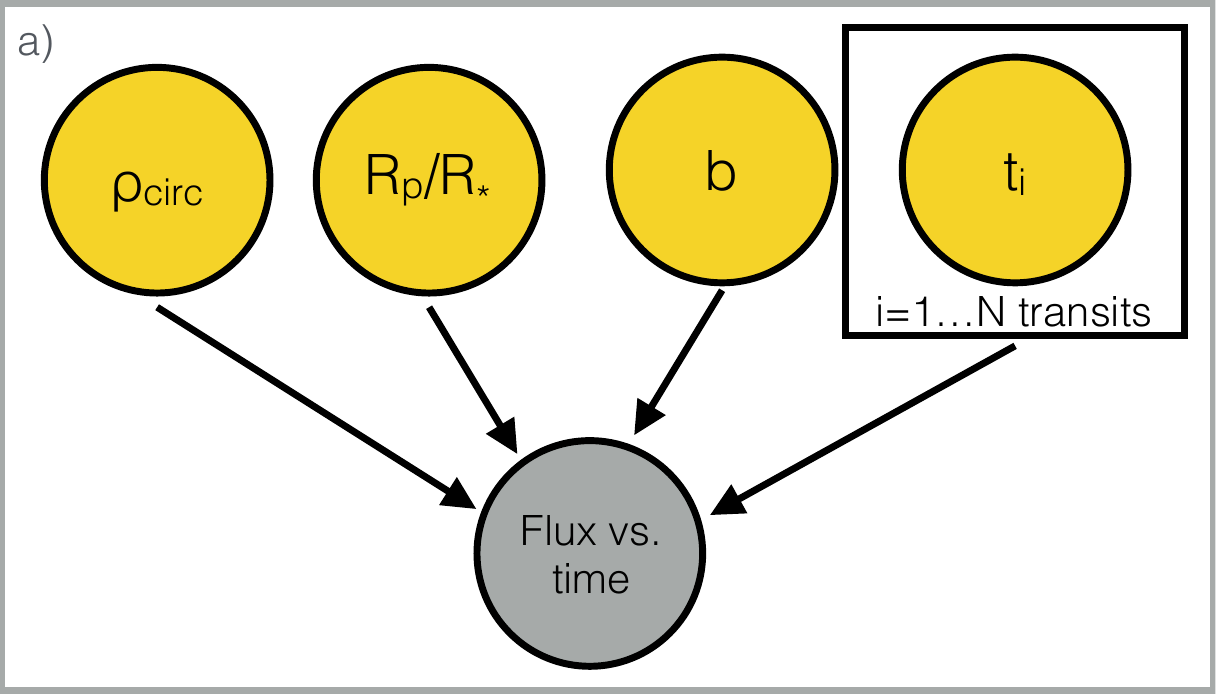}
\includegraphics[width=\columnwidth]{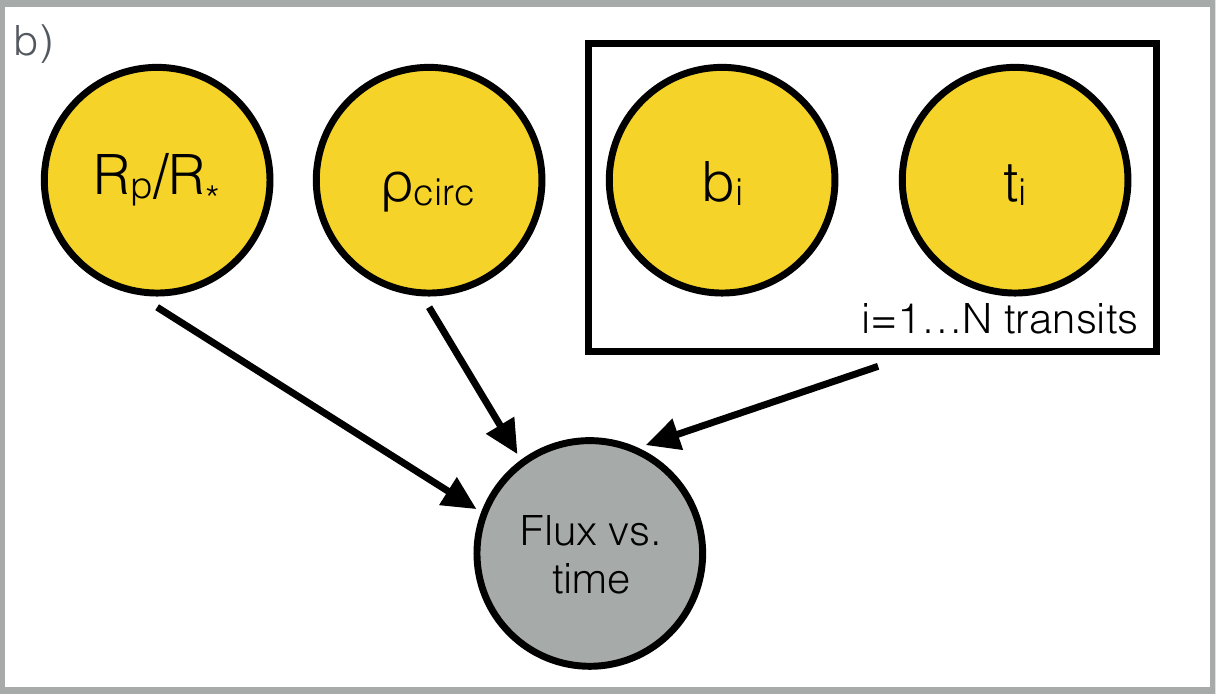}
\includegraphics[width=\columnwidth]{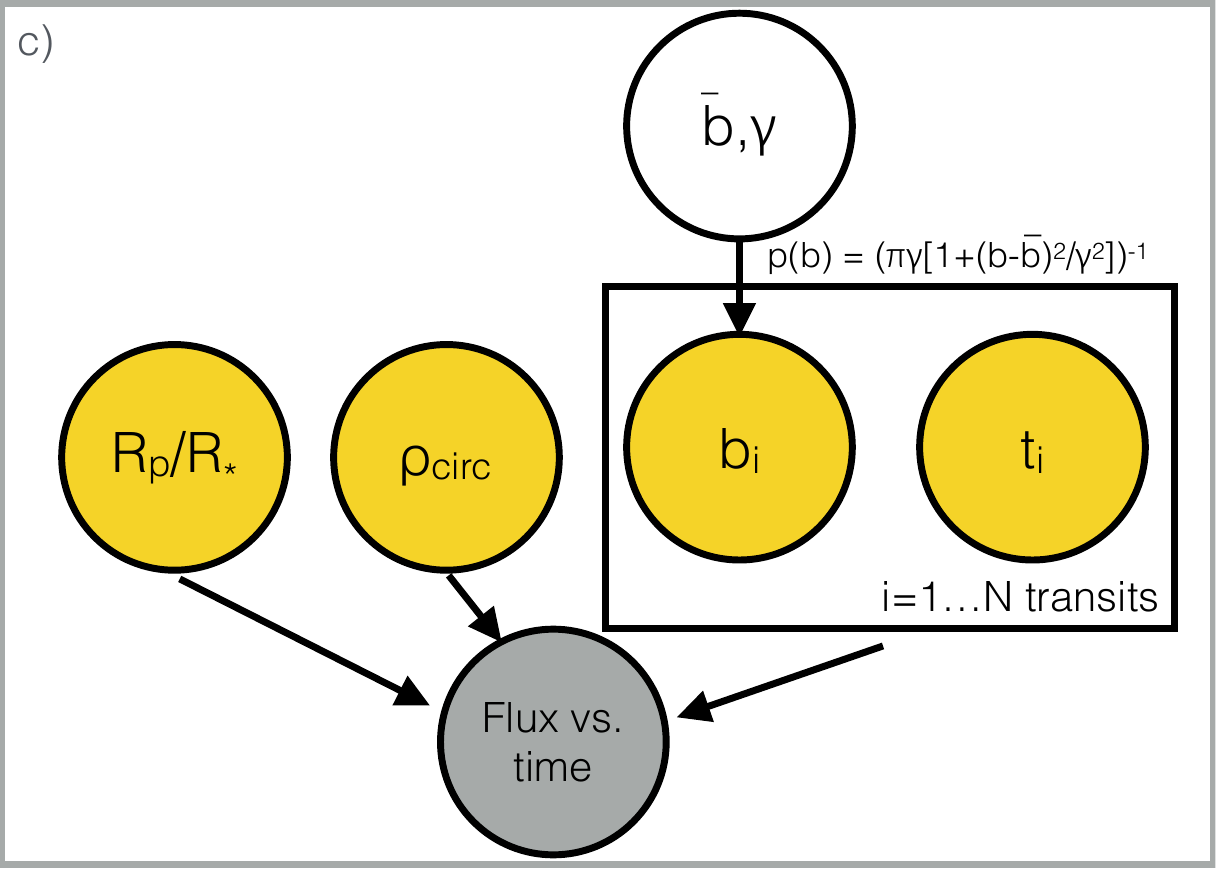}
\caption{Graphical model of inference of light curve parameters from photometric time series (flux vs. time). Yellow circles are the parameters of interest and gray the observed data. The plate (black box) indicates parameters that are individual to each ($i$) of $N$ transits. Parameters outside the plate are the same across all transits. Top (a): Impact parameter modeled as constant from transit to transit. Middle (b): Impact parameter allowed to vary from transit to transit. Bottom (c): Same as middle but with a non-uniform prior on the magnitude of change in impact parameter. The prior is a Cauchy distribution with mean impact parameter $\bar{b}$ and scale $\gamma$ of the change in impact parameter.
\label{fig:gra}
}
\end{figure}

We deduce the properties of a transiting planet based on the shape, depth, and duration of its transits. Figure \ref{fig:gra} displays graphical models of the inference of the light curve parameters from a photometric time series. The planet-to-star radius ratio, $R_p/R_\star$, sets the transit depth and affects the duration of the ingress and egress, the intervals when the planet is entering or leaving the face of the star. Each $i$ of $N$ transits has a central transit time, $t_i$. The average interval between consecutive transits is the orbital period $P$. Transit timing variations (TTVs) are deviations in the interval between transits from $P$. The impact parameter, $b$, is the scale-free distance of the transit chord from the center of the star (Fig. \ref{fig:cart}). An impact parameter $b=0$ corresponds to a transit across the stellar diameter and $b=1$ to a transit across the edge of the star. The model in which $b$ is the same from transit to transit is depicted in the top panel (a) of Figure \ref{fig:gra}. The other light curve parameter depicted in Figure \ref{fig:gra}, $\rhocirc$, relates to the transit speed. As we mentioned in Section 1, the transit speed can also be parametrized as the planet-star separation or the light curve stellar density. Here we use the latter parameter, which we denote as $\rhocirc$, the light curve stellar density assuming a circular orbit. (If the orbit is elliptical, $\rhocirc$ derived from the light curve will differ from the true stellar density.) A transit model may have additional parameters that describe the stellar limb darkening and dilution by another star in the aperture, which we will consider in later sections. See \citealt{winn10} for a detailed pedagogical treatment of transit geometry and parameters, including equations 
{ relating} $\rhocirc$ to the transit duration. We use the \citet{mand02} transit light curve model with the \citet{kipp13} limb darkening parameters. We convert our $\rhocirc$ to the \citet{mand02} normalized planet-star separation $d/R_\star$ as $${ \frac{d}{R_\star} = \left[\frac{\rhocirc}{\rho_\odot} \left(\frac{P}{P_\oplus}\right)^2\right]^{1/3} \frac{\rm au}{R_\odot}}$$ where $P_\oplus$ is the Earth's orbital period, $\rho_\odot$ is the mean solar density, and $R_\odot$ is the Sun's radius. { We employ a uniform prior on the limb darkening parameters, $P$, $b$, and $R_p/R_\star$. We use a log uniform prior on $\rhocirc$, because it is uninformative, because stellar densities themselves span many orders of magnitude, and because $\rhocirc$ can differ from $\rho_\star$ by orders of magnitude if the planet's orbit is elliptical. Moreover, we find the results are not sensitive to whether we use a uniform or log-uniform prior on $\rhocirc$. We implement this prior by fitting $\log \rhocirc$ instead of $\rhocirc$ (but report value for $\rhocirc$). Except where otherwise noted, we use the publicly available \kep  simple aperture photometry from the the Mikulski Archive for Space Telescopes (MAST).} 

\subsection{Demonstration of incorrect inference}
\label{subsec:demo}

Allowing the impact parameter to vary uniformly and independently from transit to transit (Fig. \ref{fig:gra}, panel b) results in incorrect inferences. To demonstrate the problem, we inject transits in the out-of-transit data of Kepler-419 and fit the transits with a modified version of \citet{gaza12}'s {\tt TAP} with the \citet{cart09} wavelet likelihood function. Our parameters are the planet-to-star radius ratio, the light curve stellar density, two quadratic limb darkening coefficients, a linear trend for each light curve, and white and red noise parameters for long and short cadence data. { We employ uniform priors on each linear trend's slope and intercept and on the white and red noise parameters.} See \citealt{daws15} for details of our modifications to {\tt TAP}. 

\begin{figure}
\includegraphics[width=\columnwidth]{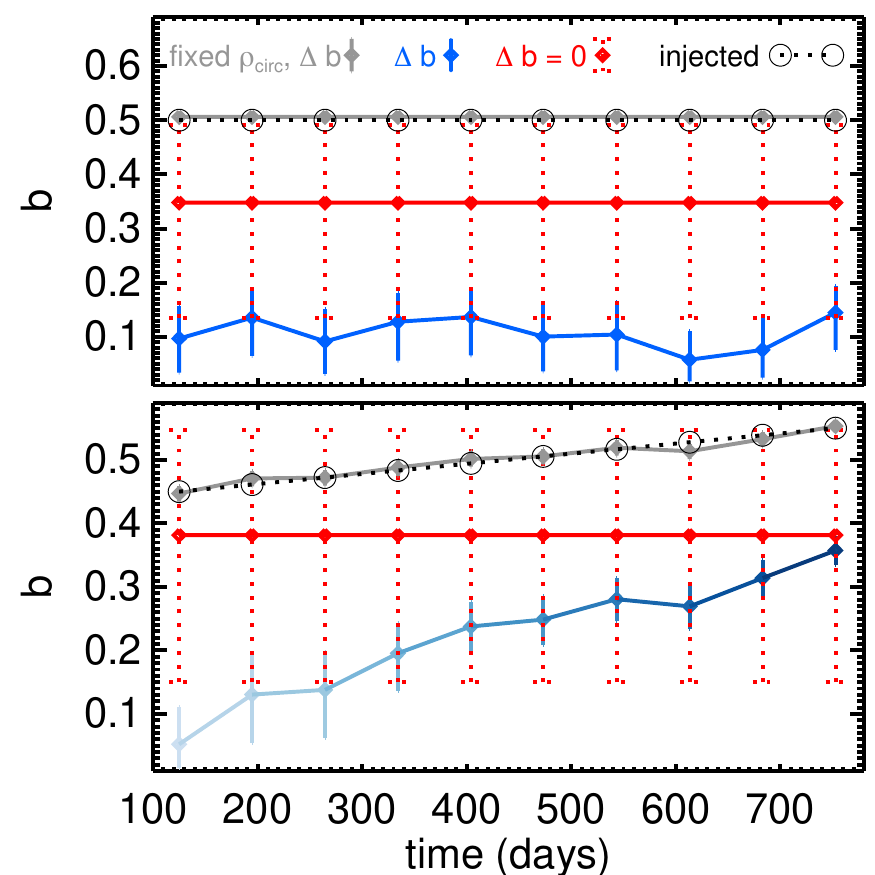}
\caption{Impact parameter vs. time injected (dashed line, circles) and recovered (diamonds with error bars, representing the median and 68\% credible interval). Top: constant injected impact parameter; bottom: changing injected impact parameter. When the impact parameter is allowed to change from transit to transit in the model (blue; panel b of Figure \ref{fig:gra}), the injected impact parameter is not recovered and the inferred change (bottom) in impact parameter is too large. When the impact parameter is modeled as constant from transit to transit (red; panel a of Figure \ref{fig:gra}), the recovered values are consistent with those injected but the change in impact parameter is by construction not detectable. When $\rhocirc$ is fixed to its true value (gray), the injected impact parameter is recovered precisely, demonstrating that the problem arises from the covariance (e.g., \citealt{cart08}) of $b$ and $\rhocirc$.
\label{fig:b}
}

\end{figure}

\begin{figure}
\begin{center}
\includegraphics[width=\columnwidth]{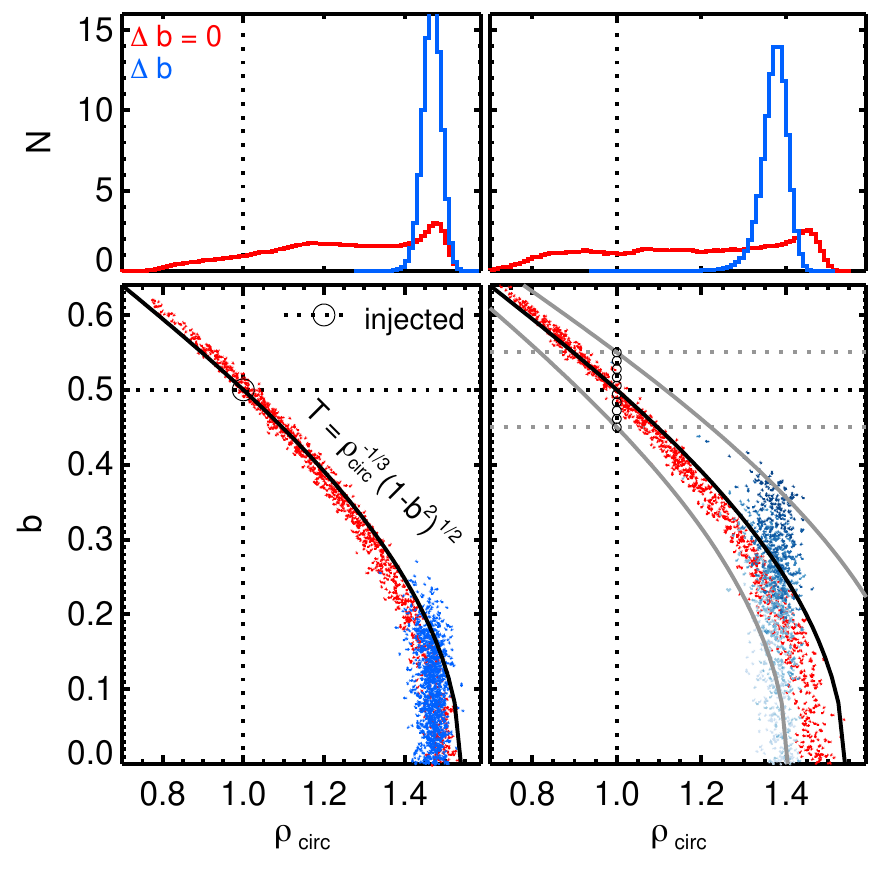}
\caption{Top: marginal posterior distribution for $\rhocirc$ when the impact parameter is modeled as constant from transit to transit (red) or allowed to vary (blue). Bottom: Two dimensional posterior distribution for $b$ vs. $\rhocirc$. Dotted lines: true injected values. Left: constant injected parameter; right: changing injected impact parameter. The solid black line (left and right) and solid gray lines (right) are the expected degeneracy between $b$ and $\rhocirc$ from a measurement of the total transit duration $T$ (Eqn. \ref{eqn:time}). When the impact parameter is allowed to vary (blue), the resulting posteriors are inconsistent with the injected values (dotted lines). 
\label{fig:brh}
}
\end{center}
\end{figure}

In the first demonstration, we inject ten transits each with a true impact parameter of $b=0.5$ (Fig. \ref{fig:b}, top panel; Fig \ref{fig:brh}, left panel). When we use the model depicted in panel a of Fig. \ref{fig:gra} that assumes the impact parameter is the same in each transit, our recovered values for the impact parameter (red; Fig \ref{fig:b}) are consistent with those injected (black circles). The two-dimensional posterior of $(b,\rhocirc)$ and marginal posterior $\rhocirc$ encompass the truth (Fig. \ref{fig:brh}). However, when we use the model depicted in panel b of Fig. \ref{fig:gra}, in which the impact parameter can vary from transit to transit, our recovered impact parameters (blue, Fig. \ref{fig:b}) are inconsistently low. The two dimensional posterior of $(b,\rhocirc)$ and marginal posterior $\rhocirc$ { exclude the truth   (Fig. \ref{fig:brh}; i.e., the true, injected values lie outside the 99.9999\% credible interval)}. When we fix $\rhocirc$ to its true value and fit each $b$, we recover the injected impact parameters (Fig. \ref{fig:b}, gray); therefore the problem arises from the covariance of $\rhocirc$ and $b$.

In the second demonstration,  we inject ten transits in which the true impact parameter varies linearly from $b=0.45$ to $b=0.55$ (Fig. \ref{fig:b}, bottom panel; Fig \ref{fig:brh}, right panel). A model that assumes $b$ is constant (red) recovers values consistent with the truth to within the uncertainties (but by construction does not capture the change). A model with $\rhocirc$ fixed to its true value (gray) recovers the inject impact parameters precisely. However, the model that allows the impact parameter to vary from transit to transit (blue) leads to inferred impact parameters that are inconsistently low and, more importantly, overestimate the change in impact parameter (Fig. \ref{fig:b}). The latter would lead to incorrect inferences about the perturber mass and orbit, including mutual inclination. The two dimensional posterior of $(b,\rhocirc)$ and marginal posterior $\rhocirc$ exclude the truth (Fig. \ref{fig:brh}{ ; i.e., the true, injected values lie outside the 99.9999\% credible interval}).

\begin{figure*}
\begin{center}
\includegraphics{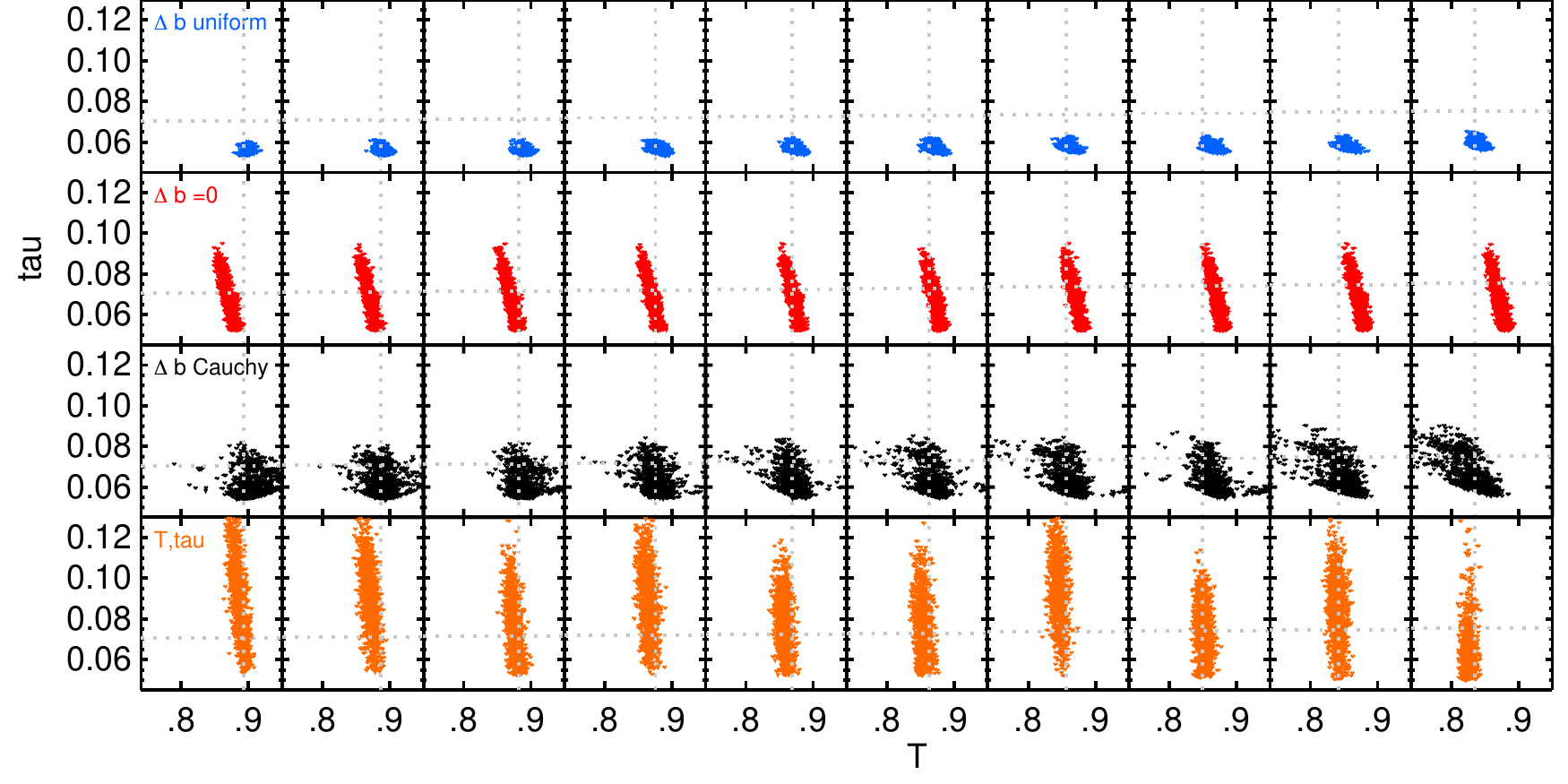}
\caption{ Two-dimensional joint posterior distribution for $T$ and $\tau$ (Eqn. \ref{eqn:time}) for each transit using full (flux vs. time) dataset for a varying b. Gray dotted lines denote the true values. When the impact parameter is allowed to vary uniformly and independent while $\rhocirc$ is the same for each transit (blue, row 1), the true values are not recovered. When the both $b$ and $\rhocirc$ are assumed to be the same for each transit (red, row 2), by definition the change in duration is not recovered (e.g., red posterior is left of the truth in first column and right of the truth in the second column. A Cauchy prior on the change in impact parameter (Section \ref{subsec:prior}; black, row 3) recovers the truth, as does fitting individual parameters to each transit with a joint prior on $\rhocirc$, $b$, and $R_p/R_\star$ that preserves a uniform prior on $T$ and $\tau$ (Section \ref{subsec:dur}; orange, row 4). Compared to the Cauchy prior approach (row 3), the individual transit approach (row 4) offers higher precision on the transit duration $T$ (but weaker precision on $\tau$) and is the best approach if one plans to directly fit the set of $T$ with a dynamical model.
\label{fig:ttau}
}
\end{center}
\end{figure*}

\begin{figure*}
\begin{center}
\includegraphics{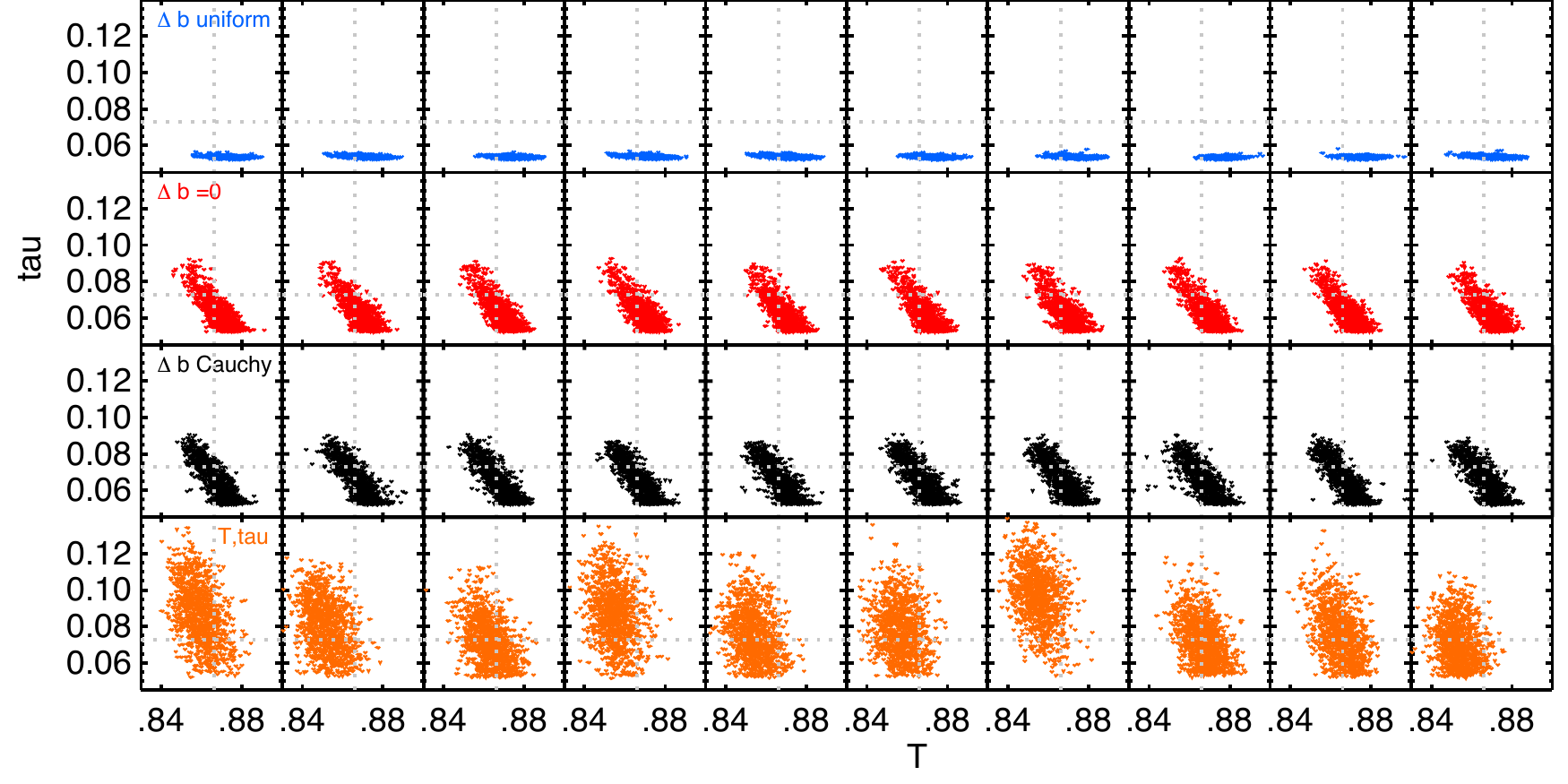}
\caption{ Two-dimensional joint posterior distribution for $T$ and $\tau$ (Eqn. \ref{eqn:time}) for each transit using full (flux vs. time) dataset for a constant b. Gray dotted lines denote the true values. When the impact parameter is allowed to vary uniformly and independent while $\rhocirc$ is the same for each transit (blue, row 1), the true values are not recovered. When the both $b$ and $\rhocirc$ are assumed to be the same for each transit (red, row 2), by definition the change in duration is not recovered (e.g., red posterior is left of the truth in first column and right of the truth in the second column. A Cauchy prior on the change in impact parameter (Section \ref{subsec:prior}; black, row 3) recovers the truth, as does fitting individual parameters to each transit with a joint prior on $\rhocirc$, $b$, and $R_p/R_\star$ that preserves a uniform prior on $T$ and $\tau$ (Section \ref{subsec:dur}; orange, row 4). 
\label{fig:ttau_nochange}
}
\end{center}
\end{figure*}

{ We also inspect the posteriors for variables corresponding to the { unit-free} full transit duration ($T_i$) and ingress/egress duration ($\tau_i$) of each ($i$) transit. We assume\footnote{In real light curves, these approximate expressions are related to the true durations by a constant in the limit where $R_p << R_\star << a$ and $|b| << 1 - R_p/R_\star$ \citep{winn10}.} { the following relations between $T,\tau,b$ and $\rhocirc$}:
\begin{align}
T= (1-b^2)^{1/2} \rhocirc^{-1/3},\nonumber \\ 
\tau = \frac{R_p}{R_\star} (1-b^2)^{-1/2}\rhocirc^{-1/3}.\nonumber \\  \label{eqn:time}
\end{align}
\noindent { where $\rhocirc$ has the units of $\rho_\odot$}. We perform inference of $b$ and $\rhocirc$ from a set of $T_i$ and $\tau_i$ using {\tt pystan} \citep{carp17,stan17}. We plot the posteriors in Fig. \ref{fig:ttau} and Fig. \ref{fig:ttau_nochange}. Because $T$ is well-constrained by the data, different treatments of $b$ lead to similar inferences. However, the model that allows the impact parameter to vary from transit to transit (blue) causes incorrect inferences of $\tau$, which is more uncertain.}

\subsection{Simplified model of light curve inference}
\label{subsec:simp}
\begin{figure}
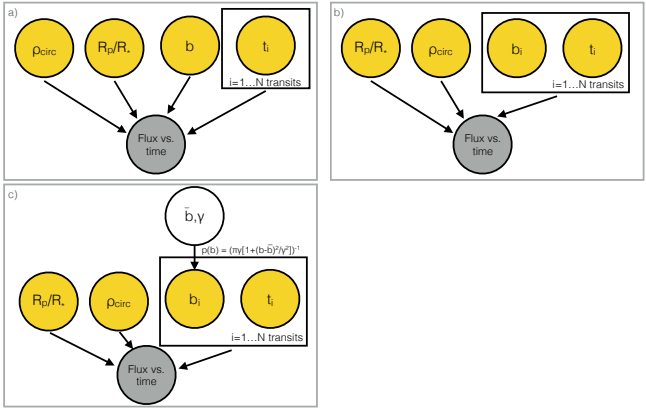

\includegraphics[width=0.49\columnwidth]{impact_link.pdf}
\includegraphics[width=0.49\columnwidth]{impact_unlink.pdf}
\includegraphics[width=0.49\columnwidth]{impact_cauchy.pdf}
\caption{Graphical model similar to Fig. \ref{fig:gra} but with light  curve parameters inferred only from a measured full transit duration ($T$) and ingress/egress duration ($\tau$) of each ($i$) transit rather than the entire transit light curve. This simplified model nonetheless reproduces the problem created by a uniform prior on the change in impact parameter. Yellow circles are the parameters of interest and gray the observed data. The plate (black box) indicates parameters that are individual to each ($i$) of $N$ transits. Parameters outside the plate are the same across all transits. Top left (a): Impact parameter modeled as constant from transit to transit. Top right (b): Impact parameter allowed to vary from transit to transit. Bottom (c): Same as middle but with a non-uniform prior on the magnitude of change in impact parameter. The prior is a Cauchy distribution with mean impact parameter $\bar{b}$ and scale $\gamma$ of the change in impact parameter.
\label{fig:gras}
}
\end{figure}

To reduce the problem demonstrated in Section \ref{subsec:demo} to its essentials, we reproduce the problem using a simplified toy model, depicted graphically in Figure \ref{fig:gras}. Instead of using the full light curve and parameter set, we use a dataset consisting of a { unit-free} full transit duration ($T_i$) and ingress/egress duration ($\tau_i$) of each ($i$) transit (Eqn. \ref{eqn:time}). Using the Stan Bayesian statistical modeling software \citep{carp17}, we fit only\footnote{The parameter $R_p/R_\star$ is also partially degenerate with $b$ and $\rhocirc$ because it affects the ingress and egress duration (e.g., \citealt{cart08}). This degeneracy makes the incorrect inference from the real dataset even more severe than in our simplified model.} the parameters $\rhocirc$ and $b$. { As with our full dataset, we use a uniform prior on $b$ and log-uniform prior on $\rhocirc$ unless otherwise noted.} The inference model with the same $b$ for each transit is shown in panel a of Fig. \ref{fig:gras} and with $b$ that can vary from transit to transit in panel b. 

\begin{figure}
\begin{center}
\includegraphics[width=3.5in]{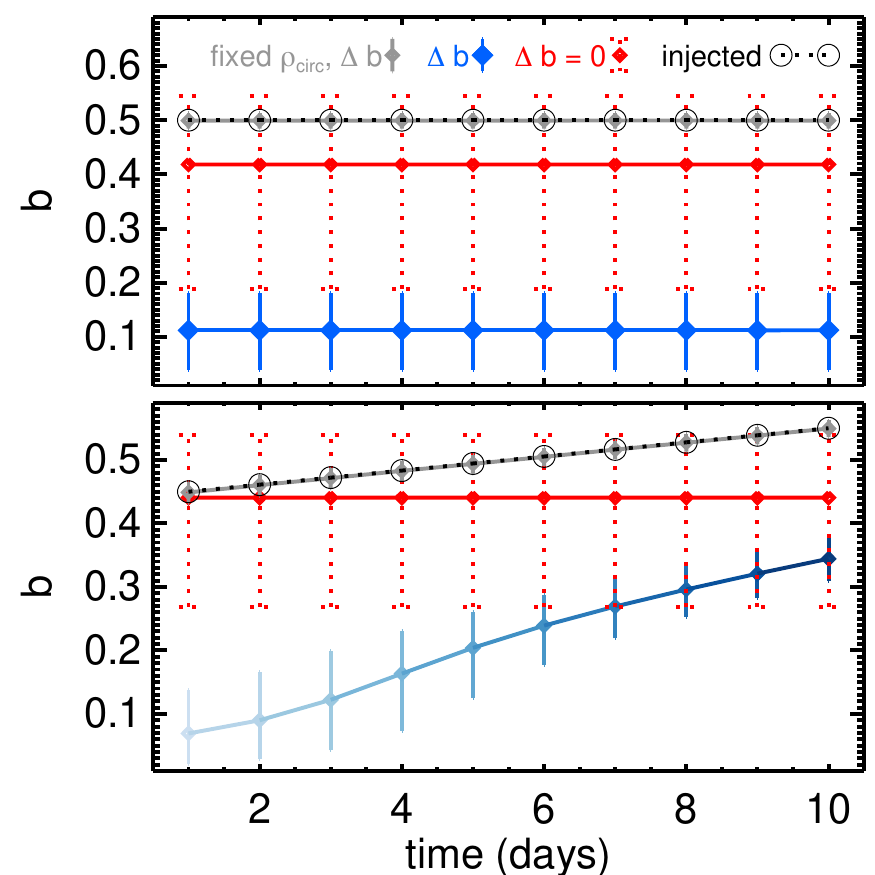}
\caption{Impact parameter vs. time injected (dashed line, circles) and recovered (diamonds with error bars, representing the median and 68\% credible interval). Same as Figure \ref{fig:b} for simplified dataset ($T, \tau$ instead of flux vs. time) depicted in Figure \ref{fig:gras}, panels a and b. The simplified dataset captures the problem: when the impact parameter is allowed to vary (blue), the injected impact parameter is not recovered and the inferred change in impact parameter is too large. Top: constant injected impact parameter; bottom: changing injected impact parameter. When the impact parameter is allowed to change from transit to transit in the model (blue; panel b of Figure \ref{fig:gras}), the injected impact parameter is not recovered and the inferred change (bottom) in impact parameter is too large. When the impact parameter is modeled as constant from transit to transit (red; panel a of Figure \ref{fig:gras}), the recovered values are consistent with those injected but the change in impact parameter is by construction not detectable. When $\rhocirc$ is fixed to its true value (gray), the injected impact parameter is recovered precisely, demonstrating that the problem arises from the covariance of $b$ and $\rhocirc$.
\label{fig:bs}
}
\end{center}
\end{figure}

\begin{figure}
\begin{center}
\includegraphics[width=\columnwidth]{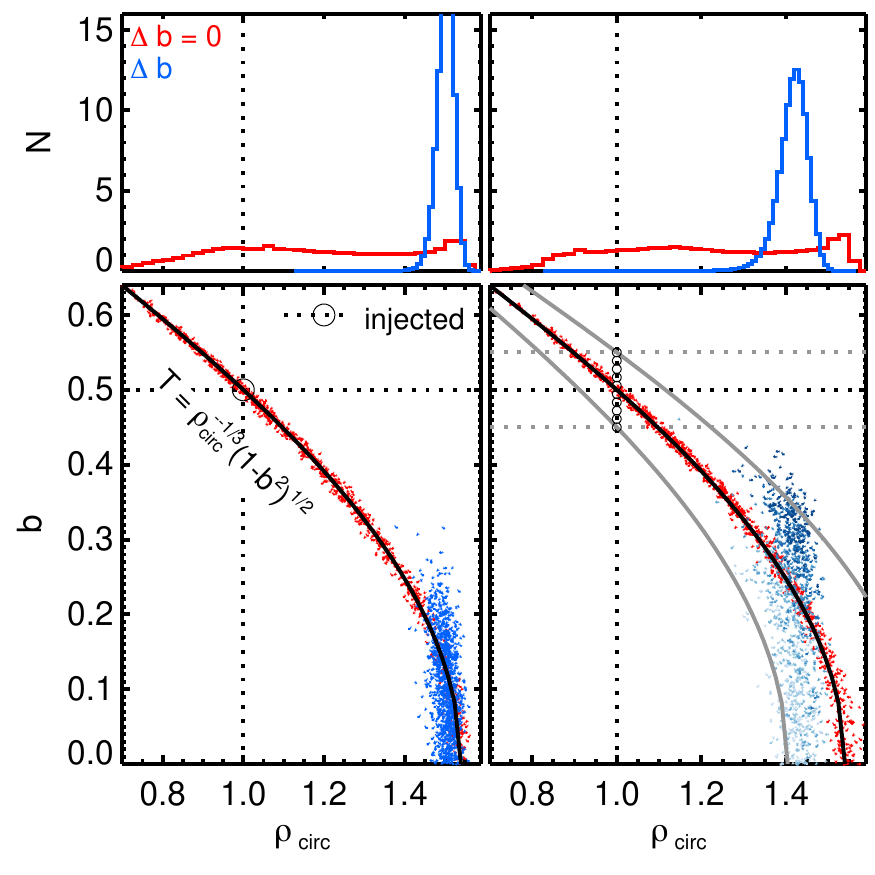}
\caption{Same as Figure \ref{fig:brh} for simplified dataset ($T, \tau$ instead of flux vs. time) depicted in Figure \ref{fig:gras}, panels a and b. The simplified dataset captures the problem: when the impact parameter is allowed to vary (blue), the resulting posteriors are inconsistent with the injected values (dotted lines). 
Top: marginal posterior distribution for $\rhocirc$ when the impact parameter is modeled as constant from transit to transit (red) or allowed to vary (blue). Bottom: Two dimensional posterior distribution for $b$ vs. $\rhocirc$. Dotted lines: true injected values. Left: constant injected parameter; right: changing injected impact parameter. The solid black line (left and right) and solid gray lines (right) are the expected degeneracy between $b$ and $\rhocirc$ from a measurement of the total transit duration $T$ (Eqn. \ref{eqn:time}).
\label{fig:brgs}
}
\end{center}
\end{figure}

In our first demonstration, we set $b=0.5$ and $\rhocirc = 1$ for each transit, compute $T$ and $\tau$, and assign each transit's $T_i$ and $\tau_i$ an uncertainty of $\sigma_T = 0.04 T$ { and $ \sigma_\tau = 0.16 \tau$} respectively. The results, shown in the top panel of Fig. \ref{fig:bs} and left panel of \ref{fig:brgs}, are very similar to full light curve inference in Fig. \ref{fig:b} and \ref{fig:brh}, demonstrating that our toy problem has captured the fundamental issue. A second demonstration, in which $b$ varies linearly from $0.45$ to $0.55$, is shown in the bottom panel of Fig. \ref{fig:bs} and right panel of \ref{fig:brgs} and also captures the problem.

\subsection{Cause of incorrect inference from transit duration variations}
\label{subsec:caus}

\begin{figure*}
\begin{center}
\includegraphics[width=\columnwidth]{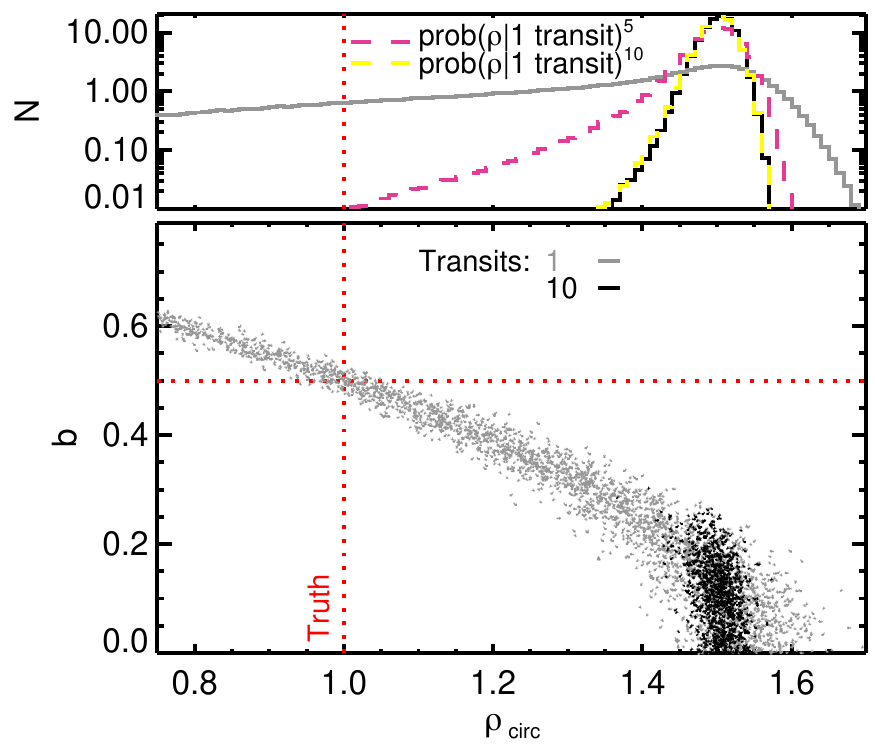}
\includegraphics[width=\columnwidth]{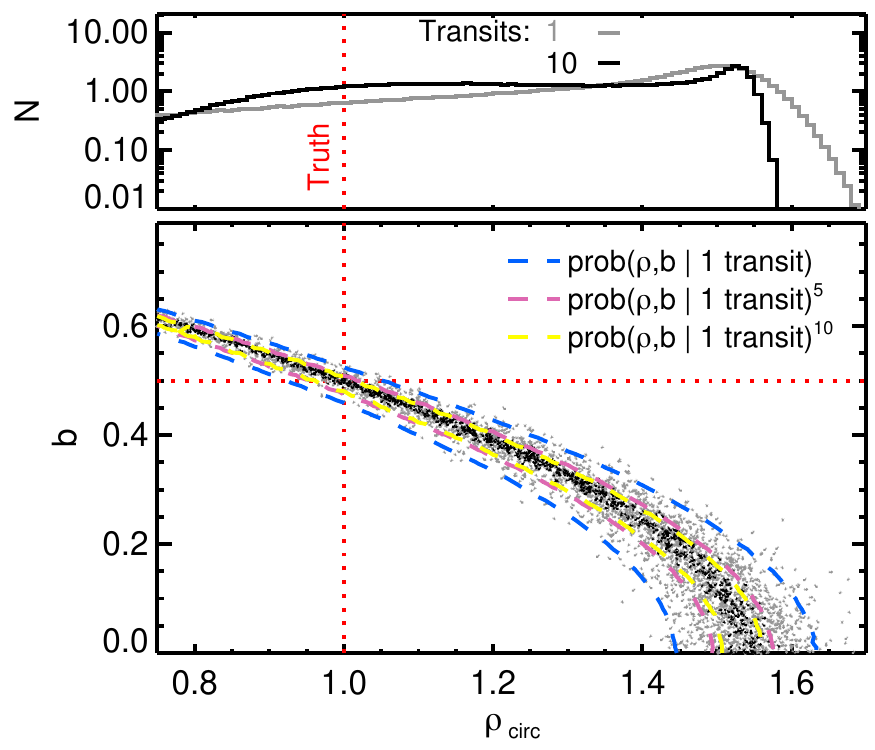}
\caption{Allowing each impact parameter to vary uniformly and independently shifts the posterior away from the truth { ($\rhocirc = 1 \rho_\odot$ and constant $b=0.5$ for each transit)} as more transits are added. Left panel: each impact parameter can vary uniformly and independently (Fig. \ref{fig:gras}, panel b). Right panel: impact parameter modeled as constant from transit to transit (Fig. \ref{fig:gras}, panel b). In both panels, gray corresponds to the inference from a single transit for the marginal $\rhocirc$ posterior (top) and joint $(\rhocirc,b)$ posterior (bottom). Black corresponds to ten transits. In the left panel, the ten transit posterior is far from the truth. In the right panel, the ten transit posterior is more accurate and precise than the one transit posterior. (Note: the model in the right panel, by construction, cannot capture a change in impact parameter.)
\label{fig:adds}
}
\end{center}
\end{figure*}

{ In the single transit case, the mode in $\rhocirc$ is not at the truth, but the $\rhocirc$ posterior includes the truth. The parameters $b$ and $\rhocirc$ are covariant (bottom panel of Fig. \ref{fig:adds}) because they both affect the transit duration $T$ (Eqn. \ref{eqn:time}). (See \citet{cart08} for a detailed exploration of their covariance.) Even though we can break the degeneracy between $b$ and $\rhocirc$ by measuring $\tau$, $\tau$ is less precisely constrained than $T$ because the ingress/egress is shorter and shallower than the full duration. For a given $\rhocirc$, the { skewed} shape of the $\rhocirc$ vs. $b$ covariance corresponds to much more posterior area at low $b$ than a high. Higher values of $\rhocirc$ correspond to larger range of $b$ consistent with the observed duration.  Incorrect inferences arise when there are multiple transits, each transit is allowed to have its own impact parameter, { and $\rhocirc$ is} constant from transit to transit.}

The simplified toy model in Section \ref{subsec:simp} elucidates the cause of the incorrect inference. The left panel of Figure \ref{fig:adds} shows how the posteriors shift away from the truth as we add more and more transits to our dataset. In the top panel, we plot the marginal posterior of $\rhocirc$. With just one transit, the true $\rhocirc$ and $b$ { ($\rhocirc = 1 \rho_\odot$ and constant $b=0.5$ for each transit)} have high probability in our posterior. Adding more transits is equivalent to raising the marginal $\rhocirc$ to the power of the number of transits (yellow dashed line): { because of the skewed shape,} the mode increases and the posterior shifts away from the truth. The right panel shows the same exercise but with $b$ assumed to be constant from transit to transit. In this case, adding more transits gets us closer to the truth. (Of course, to identify mutually inclined perturbers, we do not want to assume $b$ is constant.)

In the simplified case (Section \ref{subsec:simp}) of $N$ transits each with an identical measured $T$ and $\tau$, the marginal posterior of $\rhocirc$ for the case where the impact parameter can change from transit to transit is (Fig. \ref{fig:adds}, left panel):
\begin{equation}
    \prob\left(\rhocirc | T, \tau, N\right)\propto \left[\int_0^1 \prob\left(\rhocirc,b | T, \tau\right) db\right]^N,
\end{equation}
whereas in the case where $b$ is constant from transit to transit (Fig. \ref{fig:adds} right panel):
\begin{equation}
    \prob\left(\rhocirc | T, \tau, N\right) \propto \int_0^1 \prob\left(\rhocirc,b | T, \tau\right)^N db.
\end{equation}
The relationship between the $N$ transit posterior and one transit posterior in each case is overplotted in Fig. \ref{fig:adds}.

The marginal posterior of $b_i$ for the case where the impact parameter can change from transit to transit is:
\begin{eqnarray}
    \prob\left(b_i| T, \tau, N\right) \propto \int_0^\infty \prob\left(b_i | \tau, T, \rhocirc \right) \nonumber \\
    \times \left[\int_0^1 \prob \left(\rhocirc,b | T,\tau\right)db\right]^{N-1} d\rhocirc
\end{eqnarray}
whereas in the case where $b$ is constant from transit to transit
\begin{equation}
    \prob\left(b| T, \tau, N\right) \propto \int_0^\infty \prob\left(\rhocirc,b | T, \tau\right)^N d\rhocirc.
\end{equation}
{ Note that the proportionalities in Equations 2--5 do not include the priors on $b$ or $\rhocirc$.}

The problem arises from { how our assumptions interplay with the skewed shape of the ($\rhocirc, b$) posterior}. If we expected $b$ to truly be independent from transit to transit (if the universe randomly drew a $b$ from 0 and 1 each time the same planet transited), it would indeed be more likely for us to see  small variations in transit duration from a relatively wide range of low $b$ than from a relatively narrow range of high $b$. A uniform prior is implicitly assuming a special typical scale for the change, $\Delta b \sim 1$. In reality, favoring this special scale is not in line with the { expected} variations in impact parameter: rather, the { expected} scale of the change\footnote{We clarify that a uniform prior for the \emph{average} impact parameter is appropriate and corresponds to the reasonable assumption that other planetary systems are distributed isotropically in space.} in impact parameter spans many orders of magnitude and is typically $<< 1$. { In other words, we expect the impact parameters among different transits of the same planet to be correlated.}

\section{Mitigating the bias}
\label{sec:miti}

In the previous section, we demonstrated that incorrect inferences arise when we allow $b$ to vary { independently} from transit to transit with a uniform prior on its variation scale (while assuming $\rhocirc$ and $R_p/R_\star$ do not change detectably). { Here we present two approaches for mitigating this bias: using an appropriate prior for the change in impact parameter (Section \ref{subsec:dur}) and fitting parameters for each individual transit to identify changes in duration (Section \ref{subsec:dur}). We discuss when to use which approach and how they can be complementary in Section \ref{subsec:vs}.}

\subsection{An appropriate prior for the change in impact parameter}
\label{subsec:prior}
We argued that a uniform prior on $\Delta b$ corresponds to a favored scale for a change in $b$ that we do not truly prefer, is in fact not physically plausible, { and does not capture our expectation that impact parameters among different transits of the same planet should be correlated}. When we have no prior information about a transiting planet's perturber (or lack therefore), an uninformative prior on the \emph{scale} of the change in $b$ is most appropriate. We have found that our results are not sensitive to the functional form of the prior. One such prior that we will show works well is a Cauchy prior, which is similar to a Gaussian prior but with longer tails:
\begin{equation}
\label{eqn:cauchy}
    \prob(b) = \left[\pi \gamma \left(1+\left[b-\bar{b}\right]^2/\gamma^2\right)\right]^{-1}
\end{equation}
The likelihood function includes a product over each of $i$ impact parameters. We use a log-uniform prior for the scale $\gamma$. To capture the expected isotropic distribution of systems throughout the galaxy, we use a uniform prior on the average impact parameter $\bar{b}$. We depict this model graphically in panel c of Fig. \ref{fig:gra} and \ref{fig:gras}. 

\begin{figure}
\begin{center}
\includegraphics[width=\columnwidth]{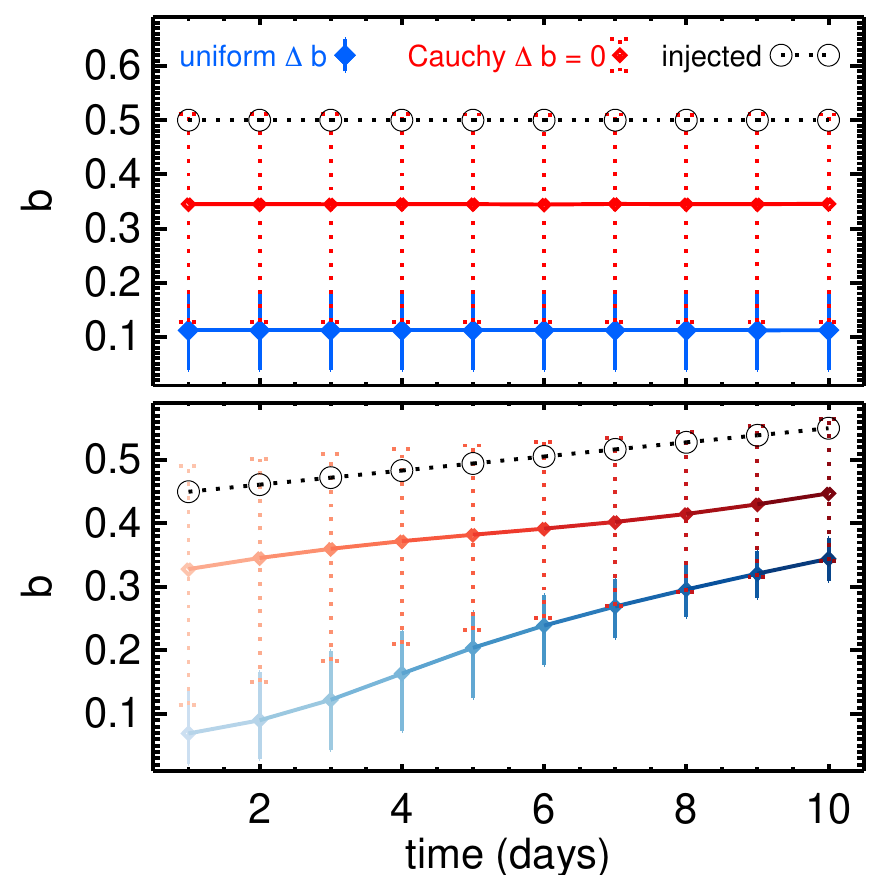}
\caption{Impact parameter vs. time injected (dashed line, circles) and recovered (diamonds with error bars, representing the median and 68\% credible interval) for simplified dataset ($T, \tau$ instead of flux vs. time) depicted in Figure \ref{fig:gras}. Top: constant injected impact parameter; bottom: changing inject impact parameter. A Cauchy prior on the change in impact parameter (red; Fig.  \ref{fig:gras}, panel c) allows us to recover values consistent with those injected, whereas a uniform prior (blue;  Fig. \ref{fig:gras}, panel b) on the change does not.
\label{fig:bsn}
}
\end{center}
\end{figure}

\begin{figure}
\begin{center}
\includegraphics[width=\columnwidth]{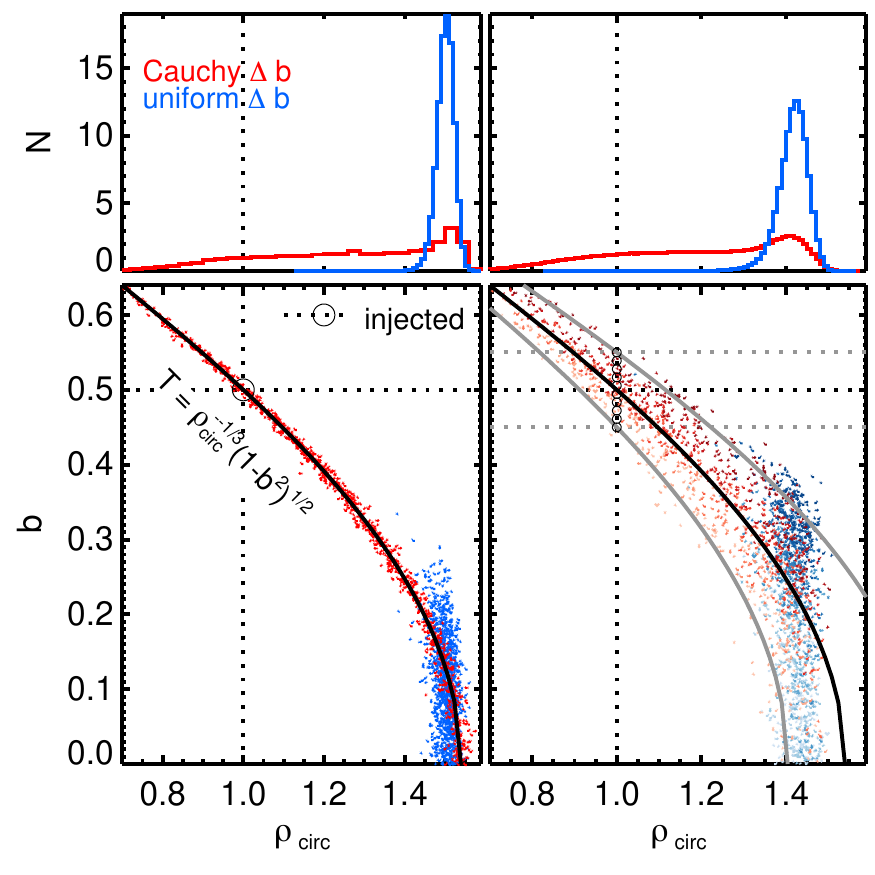}
\caption{Top: marginal posterior distribution for $\rhocirc$ when a Cauchy (red) or uniform (blue) prior is imposed on the change in impact parameter using the simplified dataset ($T, \tau$ instead of flux vs. time) depicted in Figure \ref{fig:gras}. Bottom: Two dimensional posterior distribution for $b$ vs. $\rhocirc$. Dotted lines: true injected values. Left: constant injected impact parameter; right: changing injected impact parameter. The solid black line (left and right) and solid gray lines (right) are the degeneracy between $b$ and $\rhocirc$ from a measurement of the total transit duration $T$ (Eqn. \ref{eqn:time}). The Cauchy prior (red) allows the recovery of the injected value in the posterior distribution, whereas the uniform prior (blue) does not.
\label{fig:twsn}
}
\end{center}
\end{figure}

Fig. \ref{fig:bsn} shows that this prior mitigates the problem in the simplified toy model (Fig. \ref{fig:gra}, panel c). We obtain impact parameters consistent with those injected, whether our injected $b$ is constant or varying. Fig. \ref{fig:twsn} shows that the two-dimensional $(b, \rhocirc)$ posterior and marginal $\rhocirc$ posterior encompass the truth.

\begin{figure}
\begin{center}
\includegraphics[width=\columnwidth]{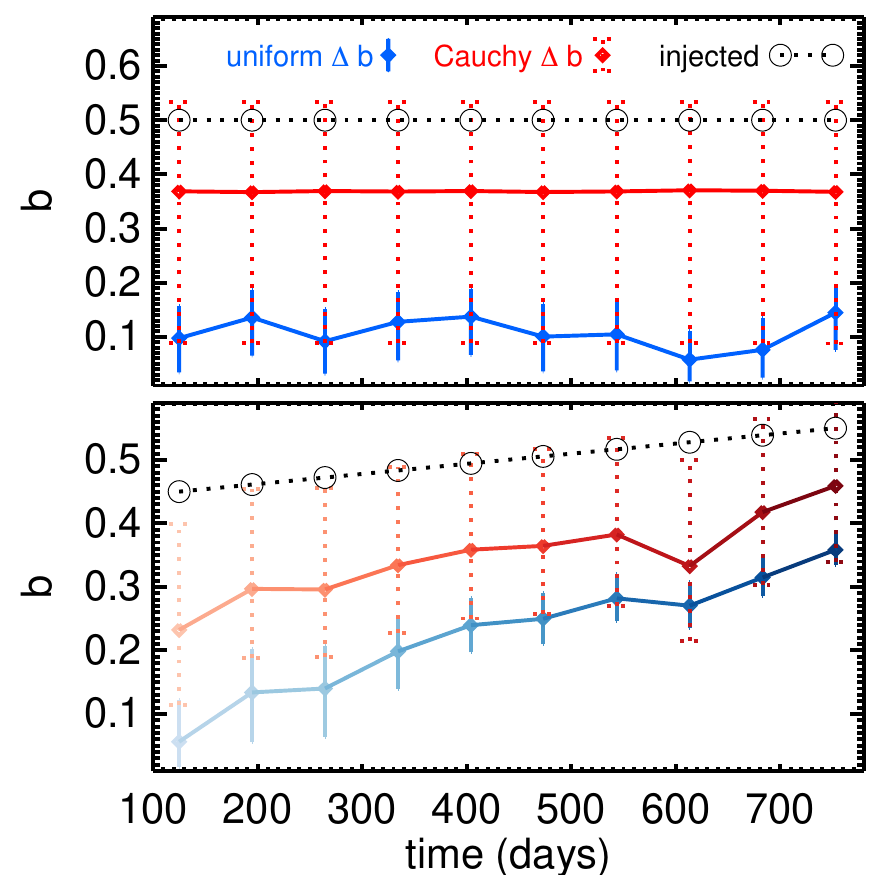}
\caption{Impact parameter vs. time injected (dashed line, circles) and recovered (diamonds with error bars, representing the median and 68\% credible interval); same as Figure \ref{fig:bsn} but using { the full (flux vs. time)} dataset. A Cauchy prior on the change in impact parameter (red; \ref{fig:gra}, panel c) allows us to recover values consistent with those injected, whereas a uniform prior (blue; Fig. \ref{fig:gra}, panel b) on the change does not. Top: constant injected impact parameter; bottom: changing inject impact parameter. 
\label{fig:bn}
}
\end{center}
\end{figure}

\begin{figure}
\begin{center}
\includegraphics[width=\columnwidth]{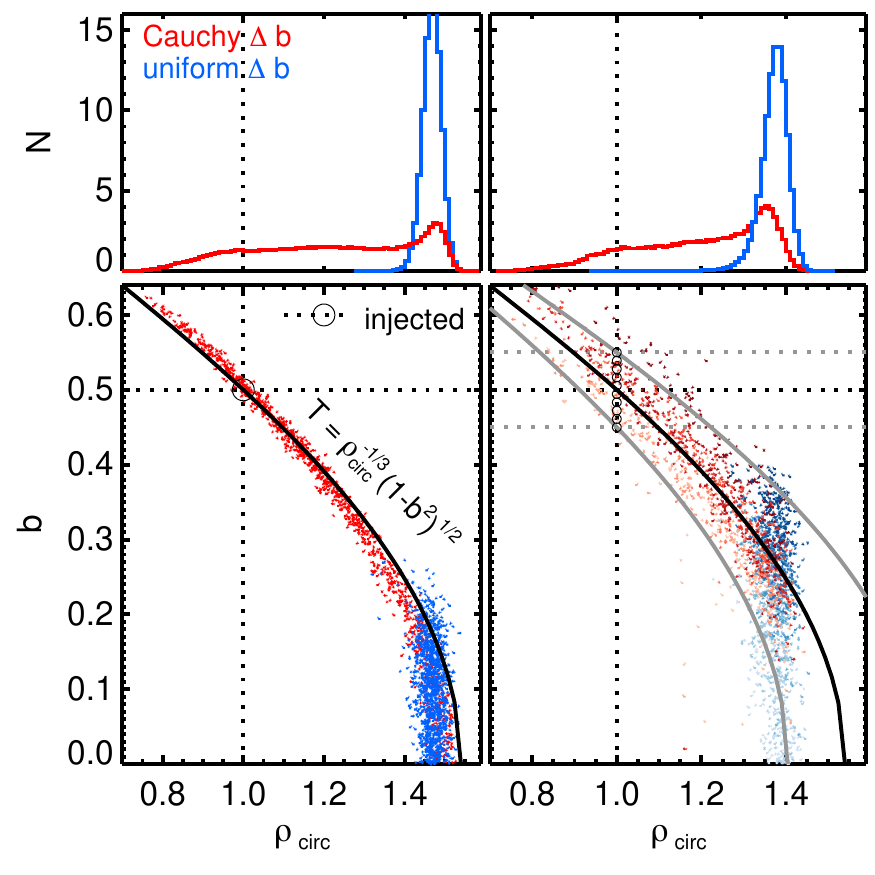}
\caption{Same as Figure \ref{fig:twsn} but using full (flux vs. { time}) dataset. A Cauchy prior on the change in impact parameter (red; \ref{fig:gra}, panel c) allows us to recover values consistent with those injected, whereas a uniform prior (blue; Fig.  \ref{fig:gra}, panel b) on the change does not. Top: marginal posterior distribution for $\rhocirc$ when a Cauchy (red) or uniform (blue) prior is imposed on the change in impact parameter. Bottom: Two dimensional posterior distribution for $b$ vs. $\rhocirc$. Dotted lines: true injected values. Left: constant injected impact parameter; right: changing injected impact parameter. The solid black line (left and right) and solid gray lines (right) are the degeneracy between $b$ and $\rhocirc$ from a measurement of the total transit duration $T$ (Eqn. \ref{eqn:time}). 
\label{fig:twn}
}
\end{center}
\end{figure}

Using this more appropriate prior also works well for full light curve fits (Fig. \ref{fig:gra}, panel c). Fig. \ref{fig:bn} and \ref{fig:twn} shows the successful recovery of parameters for transits injected into Kepler-419's out-of-transit light curve data. With the appropriate prior { on} the change in impact parameter, the posterior contains the truth for both constant and changing b. In the case of changing $b$, our truth-containing inference satisfies a prerequisite to correctly characterize the perturber causing the TDVs. We infer realistic error bars on $\rhocirc$, necessary for identifying planets on on highly elliptical orbits (e.g., \citealt{kipp10b,daws12}). With the uniform prior on the change in impact parameter (Fig. \ref{fig:twn}, blue), we might incorrectly conclude from the tight marginal distribution of $\rhocirc$ that the circular injected planet is on a moderately elliptical orbit. Our inferred values of $T$ and $\tau$ are also consistent with the truth (Fig. \ref{fig:ttau}).

\subsection{Fitting parameters for each individual transit to identify changes in duration}
\label{subsec:dur}

{ Alternatively, we can fit individual parameters to each transit to obtain robust durations and subsequently fit the collection of transit times and durations with a dynamical model. In this approach, we fit $b$, $\rhocirc$, and $R_p/R_\star$ incorporating the following term as a prior to preserve a uniform prior on the transit durations $T$ and $\tau$ (Eqn. \ref{eqn:time}) and transit depth (derived following the Appendix of \citep{burk07}):
\begin{equation}
\label{eqn:like}
    \prob\left(\rhocirc, b, R_p/R_\star\right) \propto \left(R_p/R_\star\right)^2 \frac{|b|}{1-b^2} \rhocirc^{-5/3}
\end{equation}
We caution that Eqn. \ref{eqn:like} assumes $R_p << R_\star << a$ and $|b| << 1 - R_p/R_\star$ \citep{winn10}. In the case of grazing transits, large planet-to-star radius ratio, and/or very close-in orbits, the equation must be modified.
}

{ Preserving a uniform prior on $T$, $\tau$, and depth is desirable because the dynamical model that fits inclination and eccentricity vectors will naturally impose physically realistic priors on $b$ and $\rhocirc$. (Note that the dynamical model will also need to incorporate a prior on $\rho_\star$ from a stellar model or simultaneously fit stellar parameters such as the Gaia parallax or effective temperature from the spectrum.) If we also impose priors during the light curve fit (for example, a uniform prior on $b$ and $\rhocirc$), we are applying the priors twice. However, when $T$ is well-constrained by the data -- as is typically the case for high signal-to-noise giant planet transits -- the prior on $b$ and $\rhocirc$ has a negligible effect on the inferred $T$ for each transit.}

{ An equivalent approach is to fit $T$, $\tau$, and transit depth for each light curve. In practice, we find that the above approach (fitting $b$, $\rhocirc$, and $R_p/R_\star$ with Eqn. \ref{eqn:like} as a prior) converges more quickly; in the later approach, $T$ and $\tau$ can wander off to very large values when $\tau$ is not well-constrained. Even with the above approach, we found it necessary to impose limits $-1 < b < 1$, $\rhocirc > 0$, and $R_p/R_\star > 0$ to ensure convergence.}

{ We caution that that the above approach should not be used to obtain posteriors for $\rhocirc$ and $R_p/R_\star$. These posteriors can be obtained concurrently with the dynamical model (if so, we recommend fitting the depths as part of the model) or from the approach described in Section \ref{subsec:prior}. They can be obtained less precisely by fitting a model with a joint $\rhocirc$, $b$, and $R_p/R_\star$ for all transits (Fig. \ref{fig:gra}, panel a) or fitting a binned, phase folded light curve with each transit shifted to center the mid-transit time (e.g., \citealt{masu17,vane19}).} These less precise approaches could lead to errors in $b$ and $\rhocirc$ when there are transit duration variations or, in the latter approach, large uncertainties in the TTVs that are not marginalized over \citep{kipp14}. Another approach used in the literature is to obtain an averaged { posterior} distribution by taking the median \citep{nesv14} or mean \citep{nesv12,nesv13} across transits of each posterior sample. We do not recommend using the average planet parameters from this approach, as it tends to bias the derived parameters away from the truth (Appendix, Fig. \ref{fig:indiv}). 

\subsection{Comparison of the two approaches}
\label{subsec:vs}

{ The first approach is best when the quantity of interest is the change in impact parameter, when one seeks a robust posterior for $\rhocirc$ in the presence of possible changes in impact parameter, and/or one does not plan to fit a dynamical model. The second approach is better when one seeks durations to use in a dynamical model and/or when it is unclear that changes in duration would be dominated by the change in impact parameter (a resonant system instead of a hierarchical system). As discussed in Section \ref{subsec:dur}, the second approach does not directly yield a robust posterior for $\rhocirc$, the average impact parameter, or $R_p/R_\star$.}

{ The two approaches can be complementary and used together. One can use the first approach to obtain robust posteriors for $\rhocirc$, average $b$, and $R_p/R_\star$; these quantities, along with the changes in impact parameter, can point to a good starting point for the dynamical model. The dynamical model can then be fully fit to the set of mid transit times and transit durations from the second approach.}

\section{Applications: Hierarchical Systems}
\label{sec:appl1}

We have demonstrated that allowing the impact parameter to vary uniformly and independently from transit to transit leads to incorrect inferences (Section \ref{sec:orig}). Having identified an appropriate prior on the change in impact parameter to mitigate this problem (Section \ref{sec:miti}), we will now apply this approach to systems from the literature for which changes in impact parameter or transit durations were considered in characterizing a planetary system. In this section, we will focus on hierarchical systems containing a warm Jupiter and a well-separated, non-resonant perturber that causes secular variations in the warm Jupiter's orbit. Our approach was motivated by and designed for such systems.

\subsection{Kepler-419b, a highly elliptical warm Jupiter perturbed by a non-transiting coplanar Jupiter}
\label{subsec:419}

Kepler-419b is a warm Jupiter with a 70 day orbital period on a highly elliptical ($e=0.83\pm0.01$) orbit \citep{daws12a,daws14a}. A non-transiting giant planet at 2.4 AU causes TTVs, which \citet{daws14a} used to precisely characterize the three-dimensional architecture of the system. \citet{daws14a} found from the TTVs alone that the system is coplanar, and changes in impact parameter did not offer an additional constraints. \citet{daws14a} allowed the impact parameter to vary uniformly and independently from transit to transit, which we have demonstrated leads to incorrect inferences (Section \ref{sec:orig}). Although the changes in impact parameter did not help constrain the dynamical fit, \citet{daws14a} argued that changes were detected based on the tighter constraints on $\rhocirc$ when $b$ was allowed to vary from transit to transit. Here we have shown that the tighter constraint on $\rhocirc$ is incorrect (e.g., Fig. \ref{fig:twn}).

\begin{figure}
\begin{center}
\includegraphics[width=\columnwidth]{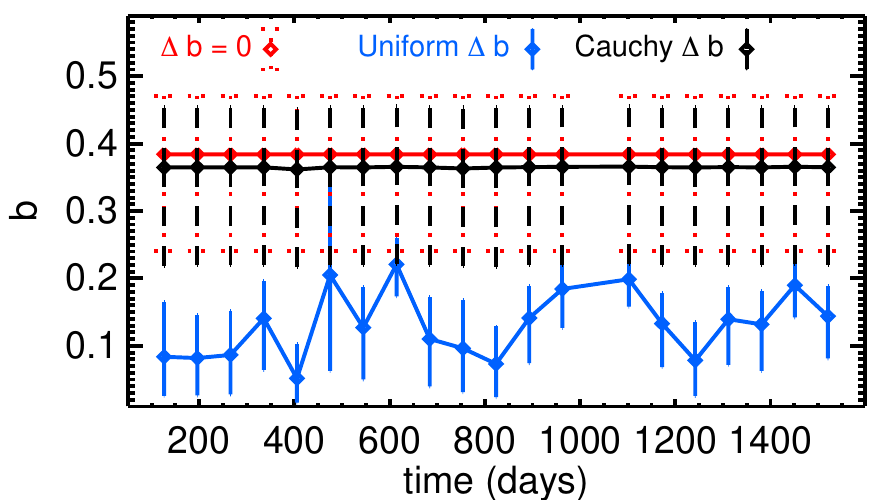}
\caption{Impact parameter vs. time fit from the Kepler-419 dataset (flux vs. time) forcing the impact parameter to be constant from transit to transit (red), allowing the impact parameter to vary with a uniform prior on the change (blue), and allowing the impact parameter to vary with a more appropriate (Section \ref{sec:miti}) Cauchy prior on the change (black).
\label{fig:419b}
}
\end{center}
\end{figure}

\begin{figure}
\begin{center}
\includegraphics[width=\columnwidth]{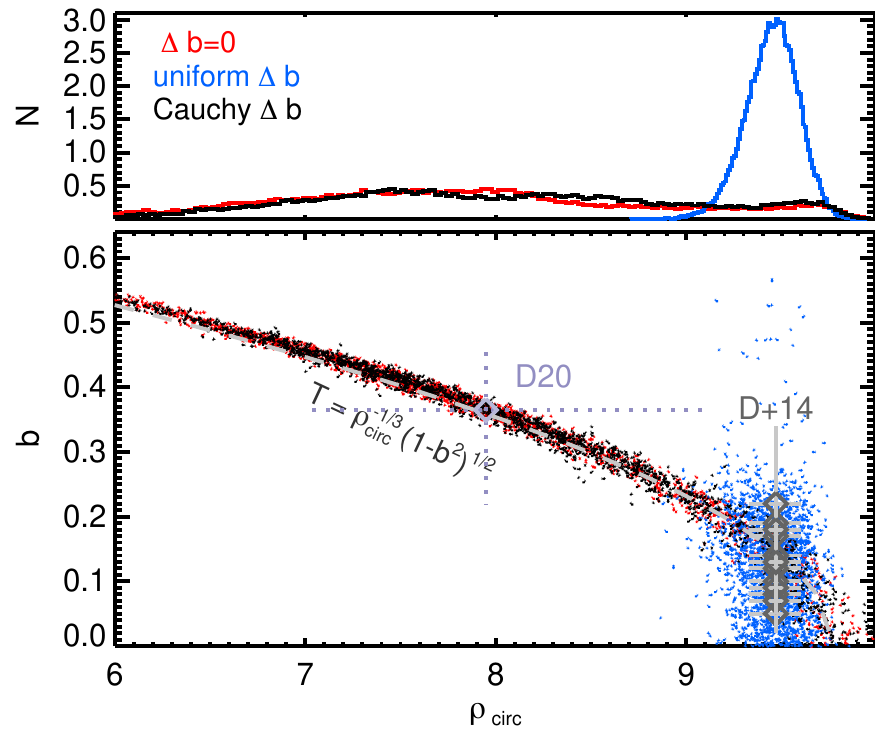}
\caption{Top: marginal posterior distribution for $\rhocirc$ when a Cauchy (red) or uniform (blue) prior is imposed on the change in impact parameter for the Kepler-419 dataset. Bottom: Two dimensional posterior distribution for $b$ vs. $\rhocirc$. D20 depicts the credible interval for the black posterior and D+14 the values reported by \citet{daws14a}.
\label{fig:419t}
}
\end{center}
\end{figure}

We perform new fits on the Kepler-419 dataset using the appropriate prior on the change in impact parameter from Section \ref{sec:miti}. We plot the impact parameter vs. time in Fig. \ref{fig:419b} and the two-dimensional posterior for $(b,\rhocirc)$ in Fig. \ref{fig:419t}. Using the Cauchy prior on the change in impact parameter (black) removes the apparent variations in impact parameter inferred from the uniform prior (blue) and also leads to a more uncertain but more realistic inference on $\rhocirc$. The results using the Cauchy prior are similar to the case where we impose $\Delta b=0$.

\begin{figure}
\begin{center}
\includegraphics[width=\columnwidth]{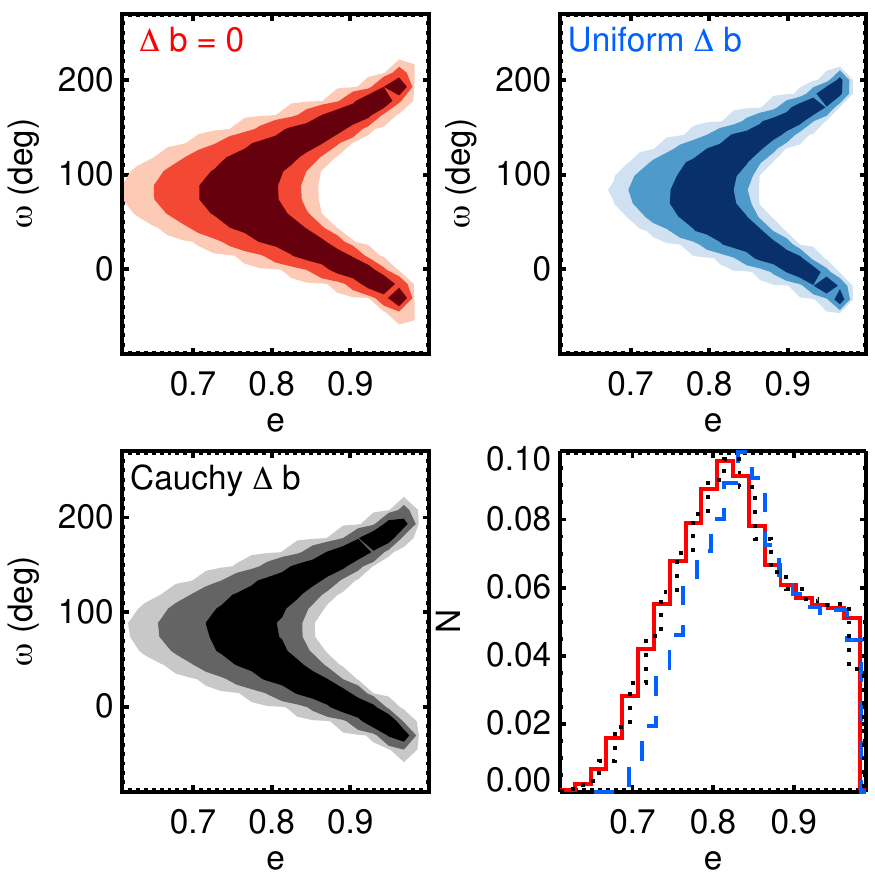}
\caption{Top left, top right, bottom left: Eccentricity vs. $\omega$ posterior distributions forcing the impact parameter to be constant from transit to transit (red, top left), allowing the impact parameter to vary with a uniform prior on the change (blue, top right), and allowing the impact parameter to vary with a Cauchy prior on the change (black, bottom left). Bottom right: marginal posterior distributions for eccentricity. A uniform prior on the change in impact parameter results in a slightly larger inferred eccentricity, but due to uncertainty in $\omega$ and $\rho_\star$, the difference is not very large.
\label{fig:419e}
}
\end{center}
\end{figure}

The $\rhocirc$ from the light curve can be combined with prior knowledge of the star's density to infer the planet's eccentricity. A falsely tight constraint on $\rhocirc$ can in principle translate to incorrect inferences on the eccentricity. In Fig. \ref{fig:419e}, we compare the eccentricity constraints derived from the three treatments of the impact parameter. In this case, we find that the degeneracy between the argument of periapse { and} eccentricity, as well as the uncertainty in the true stellar density, dominate the uncertainty in $e$. The inferred $e$ is not sensitive to the uncertainty on $\rhocirc$. We obtain similar values of $e = 0.83 ^{+0.10}_{-0.08}$, $e = 0.85 ^{+0.08}_{-0.07}$, and $e = 0.83 ^{+0.09}_{-0.08}$ using $\Delta b=0$, a uniform $\Delta b$, and a Cauchy prior on $\Delta b$ respectively. (Note that the $e=0.83\pm0.01$ derived by \citet{daws14a} is a tighter constraint because it also incorporates radial-velocity measurements, which confirm the high eccentricity measured using the ``photoeccentric'' effect.) 

For Kepler-419b, our new analysis does not qualitatively change the conclusions of \citet{daws14a} but leads to more accurate values for parameters and their uncertainties. We report these new parameters in Table \ref{tab:419}. Almost all the parameters are consistent with those of \citet{daws14a} to within uncertainties but the error bars are larger, particularly (as expected) for $\rhocirc$ and $R_{p}/R_{\star}$. The only major difference is in the average impact parameter, which is significantly larger than the individual impact parameters reported in \citet{daws14a}. This larger impact parameter is also expected from our new approach (e.g., Fig. \ref{fig:twn}).

\begin{deluxetable}{rrl}
\tabletypesize{\footnotesize}
\tablecaption{Planet Parameters for { Kepler-419b} Derived from the Light-curves \label{tab:419}}
\tablewidth{0pt}
\tablehead{
\colhead{Parameter}    & \colhead{Value\tablenotemark{a}}}
\startdata
\hline
\\
Planet-to-star radius ratio, $R_{p}/R_{\star}$   				&0.0636 &$\pm$ 0.0007 \\
Light curves stellar density, $\rhocirc$  [$\rho_\odot$]  					&7.9 &$^{+1.1}_{-0.9}$ \\
Average impact parameter, $\bar{b}$ 					&0.37 &$^{+0.09}_{-0.14}$ \\
{ Impact parameter change scale,} $\gamma$ ($10^{-5}$)&1.3&$^{+99}_{-1.3}$ \\
Limb darkening coefficient, $q_{1}$ 						& $0.30$& $^{+0.08}_{-0.07}$ 		\\
Limb darkening coefficient, $q_{2}$ 						& $0.30$&$^{+0.12}_{-0.09}$	 		\\
Red noise, short-cadence, $\sigma_r$	[ppm]		& 2400 & $\pm200$  		\\
White noise, short-cadence $\sigma_w$	[ppm]		& 655&$\pm 5$  		\\
Red noise, long-cadence $\sigma_r$		[ppm]	& 400&$\pm 60$  		\\
White noise, long-cadence $\sigma_w$	[ppm]		& 121&$\pm 7$  													\enddata
\tablenotetext{a}{The uncertainties represent the 68.3\% credible interval about the median of the posterior distribution.}
\end{deluxetable}

\subsection{Kepler-693b, a moderately elliptical warm Jupiter perturbed by a non-transiting, mutually inclined brown dwarf}
\label{subsec:693}
Kepler-693b is a warm Jupiter that exhibits transit timing and duration variations due to the perturbations of a non-transiting brown dwarf, Kepler-693c, hierarchically separated at several AU and with a large mutual inclination \citep{masu17}. The brown dwarf causes secular oscillations in the warm Jupiter, allowing the warm Jupiter's orbit to periodically get close enough to the star for tidal circularization. Therefore Kepler-693c is exactly the type of companion expected to a warm Jupiter achieving its short period through high eccentricity tidal migration. { \citep{masu17}'s analysis of Kepler-693b and Kepler-448b (Section \ref{subsec:448}) was not subject to the bias described in Section \ref{sec:orig}. They followed the approach described in Section \ref{subsec:dur} of fitting individual parameters to each transit to obtain transit times and durations to fit with a dynamical model.}

\begin{figure}
\begin{center}
\includegraphics[width=\columnwidth]{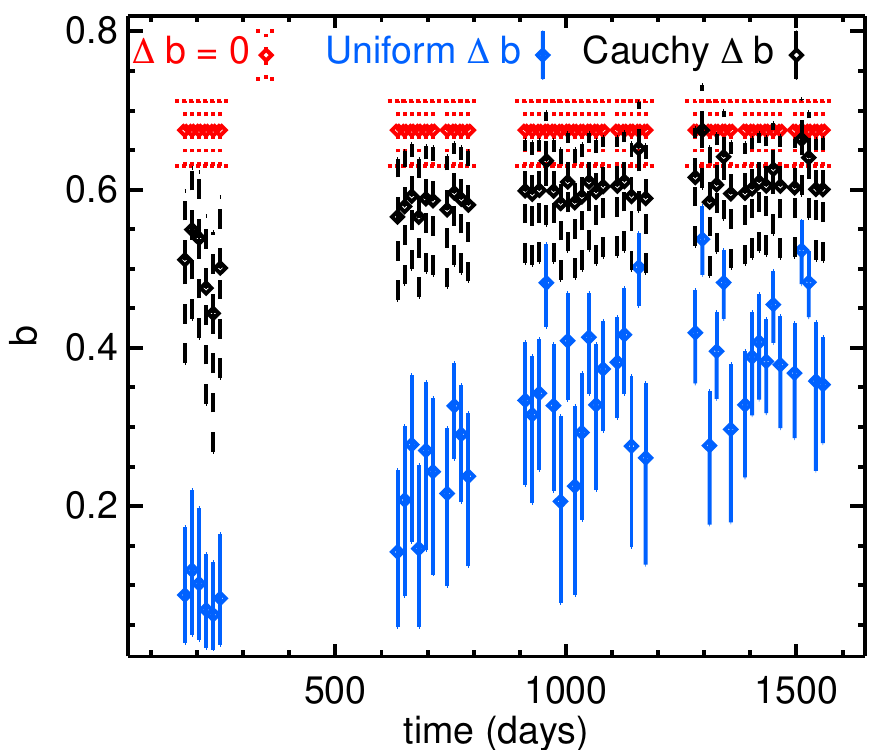}
\caption{Impact parameter vs. time fit from the Kepler-693 dataset (flux vs. time) forcing the impact parameter to be constant from transit to transit (red), allowing the impact parameter to vary with a uniform prior on the change (blue), and allowing the impact parameter to vary with a more appropriate (Section \ref{sec:miti}) Cauchy prior on the change (black). The Cauchy prior on the change in impact parameter (black) allows for the confirmation of a change in impact parameter for Kepler-693b, but this change is more modest than inferred using a uniform prior (blue).
\label{fig:693b}
}
\end{center}
\end{figure}

\begin{figure}
\begin{center}
\includegraphics[width=\columnwidth]{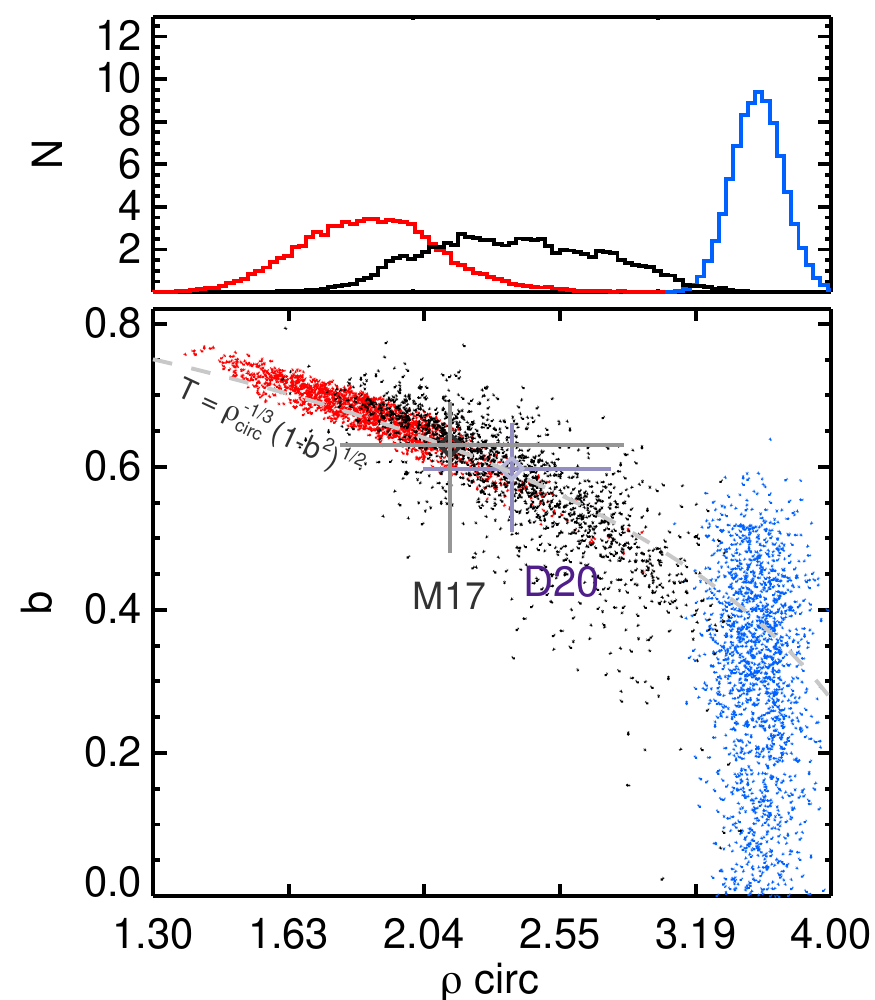}
\caption{Top: marginal posterior distribution for $\rhocirc$ when a Cauchy (red) or uniform (blue) prior is imposed on the change in impact parameter for the Kepler-693 dataset. Bottom: Two dimensional posterior distribution for $b$ vs. $\rhocirc$. Best fit values from \citet{masu17} are indicated. D20 depicts the credible interval for the black posterior.
\label{fig:693t}
}
\end{center}
\end{figure}

\begin{figure}
\begin{center}
\includegraphics[width=0.57\columnwidth]{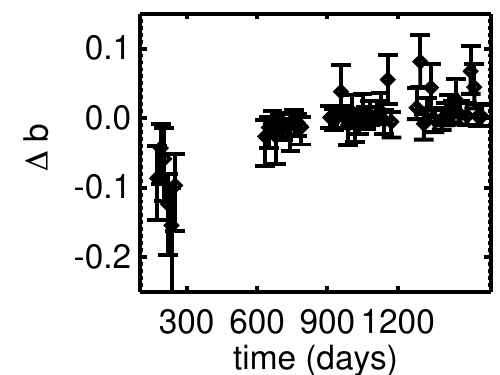}\includegraphics[width=0.43\columnwidth]{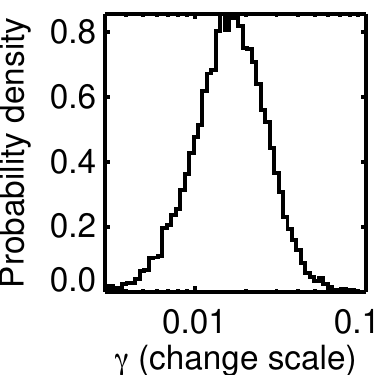}
\caption{Kepler-693b. Left: Change in impact parameter from its median value, from fit using Cauchy prior on scale for change in impact parameter. Right: Posterior for impact parameter change scale $\gamma$. We confirm that Kepler-693b exhibits a significant change in impact parameter over the Kepler Mission.
\label{fig:693g}
}
\end{center}
\end{figure}

In Fig. \ref{fig:693b} and \ref{fig:693t}, we plot the results of our light curve fits for Kepler-693. We report our best-fit parameters in Table \ref{tab:693b}. Our light curve parameters are consistent with \citet{masu17} to within the { uncertainties.} Consistent with \citet{masu17}'s TDV detections, we detect a change in impact parameter of Kepler-693b (Fig. \ref{fig:693g}). The change scale { is $\gamma=0.018^{+0.011}_{-0.007}$}. If we had allowed the impact parameter to vary uniformly and independently, we would have overestimated the magnitude of the change (Fig. \ref{fig:693b}). 

{ \citet{masu17} derived average values for transit parameters from a fit to a binned, phased-folded light curve (see Section \ref{subsec:dur} for a discussion of this approach). Our constraints on the average $\rhocirc$ and transit impact parameter (Fig. \ref{fig:693t}) are somewhat more precise.}

\begin{deluxetable}{rrl}
\tabletypesize{\footnotesize}
\tablecaption{Planet Parameters for Kepler-693b Derived from the Light-curves \label{tab:693b}}
\tablewidth{0pt}
\tablehead{
\colhead{Parameter}    & \colhead{Value\tablenotemark{a}}}
\startdata
\hline
\\
Planet-to-star radius ratio, $R_{p}/R_{\star}$   				&0.116 &$^{+0.004}_{-0.003}$\\
Light curves stellar density, $\rhocirc$  [$\rho_\odot$]  					&2.5 &$^{+0.4}_{-0.4}$ \\
Average impact parameter, $\bar{b}$ 					&0.57 &$^{+0.07}_{-0.09}$ \\
{ Impact parameter change scale,} $\gamma$ &0.018&$^{+0.011}_{-0.007}$ \\
Limb darkening coefficient, $q_{1}$ 						& $0.5$& $^{+0.3}_{-0.2}$ 		\\
Limb darkening coefficient, $q_{2}$ 						& $0.5$&$^{+0.3}_{-0.2}$	 		\\
Red noise, short-cadence, $\sigma_r$	[ppm]		& 7000 & $\pm2000$  		\\
White noise, short-cadence $\sigma_w$	[ppm]		& 6010&$^{+40}_{-30}$  		\\
Red noise, long-cadence $\sigma_r$		[ppm]	& 1300&$^{+500}_{-600}$  		\\
White noise, long-cadence $\sigma_w$	[ppm]		& 1190&$\pm 30$  													\enddata
\tablenotetext{a}{The uncertainties represent the 68.3\% credible interval about the median of the posterior distribution.}
\end{deluxetable}

\subsection{Kepler-448b, { an elliptical} warm Jupiter perturbed by a non-transiting brown dwarf}
\label{subsec:448}
\citet{masu17} also detected a non-transiting brown dwarf companion to warm Jupiter Kepler-448b using transit timing variations. \citet{masu17} found that Kepler-448b did not exhibit significant transit duration variations and that the mutual inclination of Kepler-448c is poorly constrained. Therefore it is uncertain whether secular oscillations allow Kepler-448b to get close enough to the star for tidal migration. We fit the light curves and do not detect a significant change in impact parameter (Table \ref{tab:448b}). Our light curve parameters are consistent with \citet{masu17} except for a small but significant discrepancy in the radius ratio, which may be due to different approaches for treating correlated noise. We echo \citet{masu17}'s hope that Gaia observations may shed light on the mutual inclination between Kepler-448b and c.

\begin{deluxetable}{rrl}
\tabletypesize{\footnotesize}
\tablecaption{Planet Parameters for Kepler-448b Derived from the Light-curves \label{tab:448b}}
\tablewidth{0pt}
\tablehead{
\colhead{Parameter}    & \colhead{Value\tablenotemark{a}}}
\startdata
\hline
\\
Planet-to-star radius ratio, $R_{p}/R_{\star}$   				&0.08993 &$^{+0.00007}_{-0.00008}$ \\
Light curves stellar density, $\rhocirc$  [$\rho_\odot$]  					&0.282 &$^{+0.002}_{-0.002}$ \\
Average impact parameter, $\bar{b}$ 					&0.359 &$^{+0.006}_{-0.006}$ \\
{ Impact parameter change scale,} $\gamma$ ($10^{-4}$)&1.1&$^{+5.6}_{-0.9}$ \\
Limb darkening coefficient, $q_{1}$ 						& $0.221$& $^{+0.008}_{-0.008}$ 		\\
Limb darkening coefficient, $q_{2}$ 						& $0.34$&$^{+0.02}_{-0.02}$	 		\\
Red noise, short-cadence, $\sigma_r$	[ppm]		& 3600& $\pm30$  		\\
White noise, short-cadence $\sigma_w$	[ppm]		& 247.1&$^{+0.7}_{-0.6}$  		\\												\enddata
\tablenotetext{a}{The uncertainties represent the 68.3\% credible interval about the median of the posterior distribution.}
\end{deluxetable}

\section{Applications: Near Resonant Systems}
\label{sec:appl2}

Although our approach is designed for hierarchical systems, here we explore its application to systems near orbital resonance. These systems have been more commonly characterized using transit time and duration variations than hierarchical systems. Although sometimes our assumption that the change in transit duration is dominated by a change in impact parameter does not hold, we will show that our approach is nonetheless useful for robustly identifying changes in impact parameter. 

\subsection{Kepler-46b, a warm Jupiter perturbed by a non-transiting, nearly coplanar warm Saturn: evidence for TDVs}
\label{subsec:872}
Kepler-46b, a warm Jupiter, was the first planet to have its non-transiting companion characterized without degeneracy by TTVs \citep{nesv12}. The warm Jupiter's non-transiting companion, a warm Saturn, may have small mutual inclination \citep{saad17}. To assess the TDVs, \citet{nesv12} fit the data using a model in which each transit had its own $\rhocirc$, $b$, and $R_p/R_\star$ { (Section \ref{subsec:dur}). Their analysis  was not subject to the bias described in Section \ref{sec:orig}.} They found no significant TDVs.

The lack of TDVs allowed them to rule one of two solutions that were both consistent with the TTVs. However, the transit durations did not offer a meaningful constraint on that favored solution. \citet{saad17} further refined the system's parameters using TTVs alone with a longer baseline of the full \kep dataset and found that favored solution to be a much better fit.

\begin{figure}
\begin{center}
\includegraphics[width=\columnwidth]{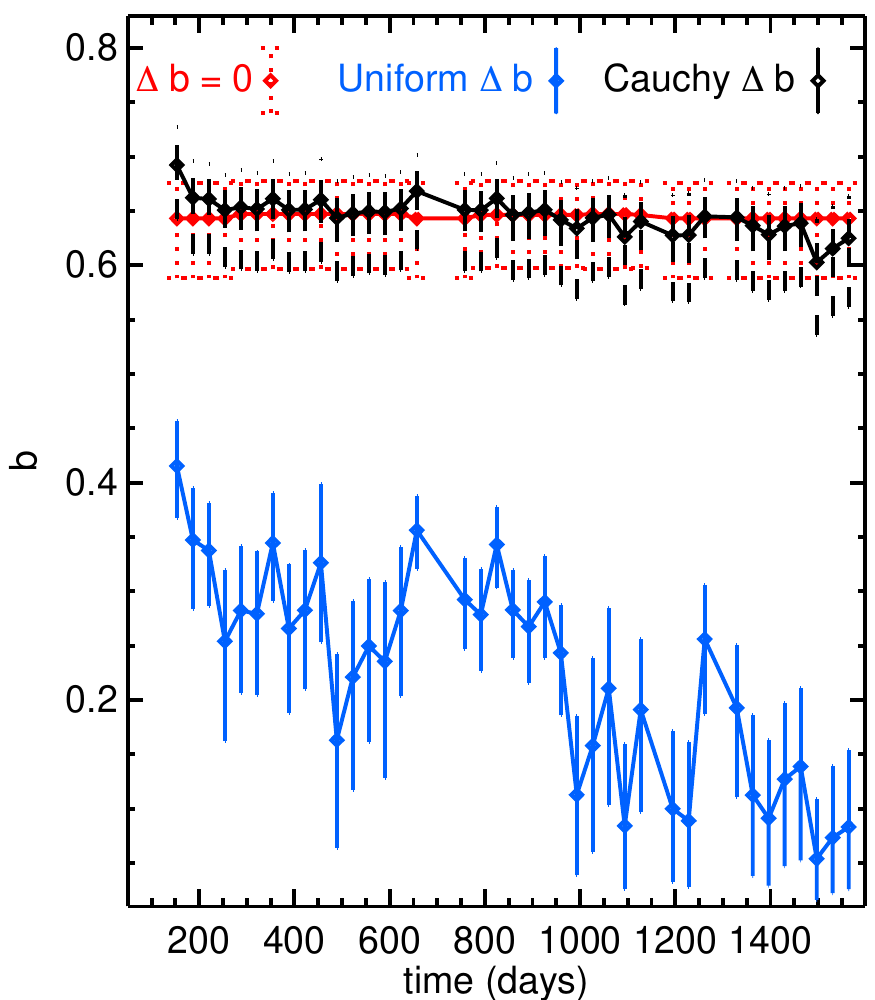}
\caption{Impact parameter vs. time fit from the Kepler-46 dataset (flux vs. time) forcing the impact parameter to be constant from transit to transit (red), allowing the impact parameter to vary with a uniform prior on the change (blue), and allowing the impact parameter to vary with a more appropriate (Section \ref{sec:miti}) Cauchy prior on the change (black). The Cauchy prior on the change in impact parameter (black) allows for the detection of a change in impact parameter for Kepler-46b, but this change is more modest than inferred using a uniform prior (blue).
\label{fig:872b}
}
\end{center}
\end{figure}

\begin{figure}
\begin{center}
\includegraphics[width=\columnwidth]{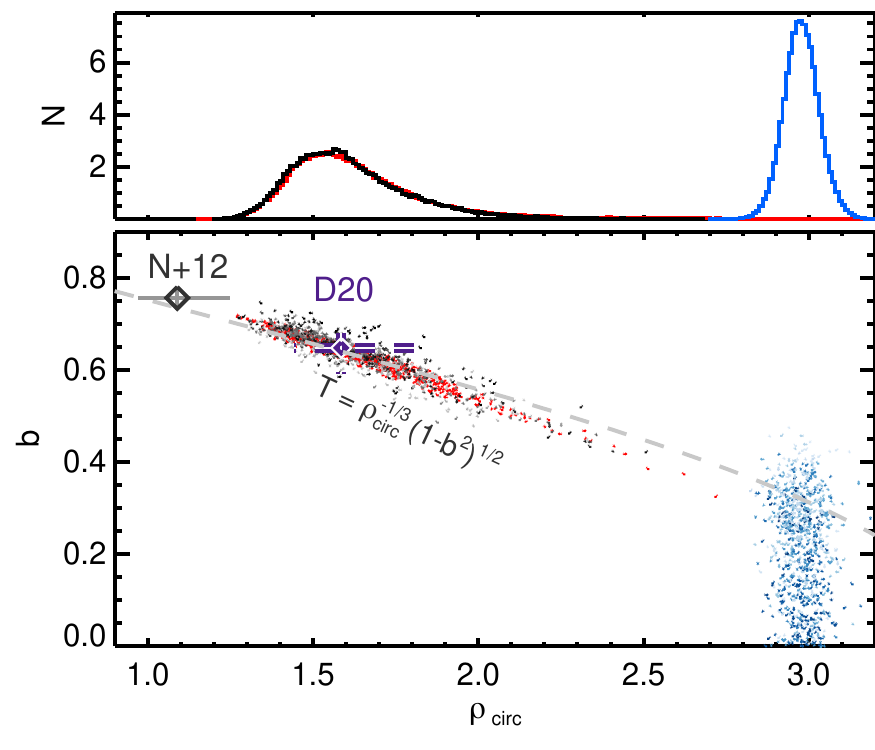}
\caption{Top: marginal posterior distribution for $\rhocirc$ when a Cauchy (red) or uniform (blue) prior is imposed on the change in impact parameter for the Kepler-46 dataset. Bottom: Two dimensional posterior distribution for $b$ vs. $\rhocirc$. Best fit values from \citet{nesv12} are indicated. D20 depicts the credible interval for the black posterior.
\label{fig:872t}
}
\end{center}
\end{figure}

\begin{figure}
\begin{center}
\includegraphics[width=0.57\columnwidth]{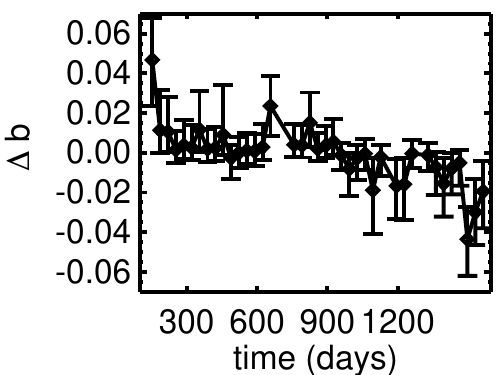}\includegraphics[width=0.43\columnwidth]{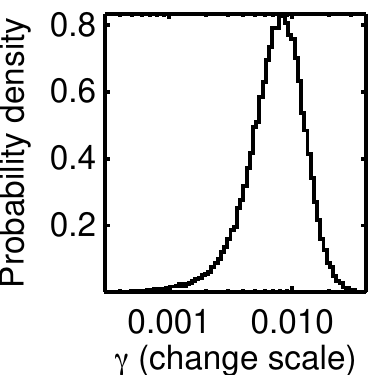}
\caption{Kepler-46b. Left: Change in impact parameter from its median value, from fit using Cauchy prior on scale for change in impact parameter. Right: Posterior for impact parameter change scale $\gamma$. We find that Kepler-46b exhibits a modest but significant change in impact parameter over the Kepler Mission.
\label{fig:872g}
}
\end{center}
\end{figure}

Following the procedure described in Section \ref{subsec:419}, we fit the full dataset and find evidence for a change in impact parameter (Fig. \ref{fig:872b}, Fig. \ref{fig:872t}, Table \ref{tab:46}). Allowing the impact parameter to vary uniformly and independently from transit to transit results in a large change in impact parameter. With an appropriate prior, the impact parameter still changes but more modestly yet still significantly. The scale for the change is $\gamma=0.008^{+0.004}_{-0.003}$. We plot the change in impact parameter and $\gamma$ posterior in Fig. \ref{fig:872g}. 

\begin{deluxetable}{rrl}
\tabletypesize{\footnotesize}
\tablecaption{Planet Parameters for Kepler-46b Derived from the Light-curves \label{tab:46}}
\tablewidth{0pt}
\tablehead{
\colhead{Parameter}    & \colhead{Value\tablenotemark{a}}}
\startdata
\hline
\\
Planet-to-star radius ratio, $R_{p}/R_{\star}$   				&0.0816 &$^{+0.0015}_{-0.0019}$ \\
Light curves stellar density, $\rhocirc$  [$\rho_\odot$]  					&1.58 &$^{+0.20}_{-0.14}$ \\
Average impact parameter, $\bar{b}$ 					&0.65 &$^{+0.03}_{-0.05}$ \\
{ Impact parameter change scale,} $\gamma$ &0.008&$^{+0.004}_{-0.003}$ \\
Limb darkening coefficient, $q_{1}$ 						& $0.46$& $^{+0.23}_{-0.13}$ 		\\
Limb darkening coefficient, $q_{2}$ 						& $0.37$&$^{+0.26}_{-0.19}$	 		\\
Red noise, short-cadence, $\sigma_r$	[ppm]		& 3100 & $\pm500$  		\\
White noise, short-cadence $\sigma_w$	[ppm]		& 1981&$\pm 7$  		\\
Red noise, long-cadence $\sigma_r$		[ppm]	& 900&$\pm150$  		\\
White noise, long-cadence $\sigma_w$	[ppm]		& 412&$\pm 14$  													\enddata
\tablenotetext{a}{The uncertainties represent the 68.3\% credible interval about the median of the posterior distribution.}
\end{deluxetable}

Our results for $R_{p}/R_{\star}$, $\rhocirc$, and $\bar{b}$ are inconsistent at several sigma with \citet{nesv12}, who find $R_{p}/R_{\star} = 0.0887^{+0.0010}_{-0.0012}$, $\rhocirc = 1.09^{+0.16}_{-0.12} \rho_\odot$, and $\bar{b} =0.757^{+0.022}_{-0.027}$. The difference in $R_{p}/R_{\star}$ may be due to the treatment of dilution from other stars in the aperture. \citet{nesv12} assumed a dilution factor based on the median of simple aperture photometry (SAP) vs. the median of the presearch data conditioned (PDC) photometry for each quarter, assuming that the latter has been corrected for dilution. \citet{nesv12} infer a larger radius ratio due to their dilution correction. However, we find that the reported crowding metric indicates that no dilution correction has been applied to the presearch data conditioned (PDC) photometry. The PDC photometry does have a different median, but we find the difference is multiplicative, rather than additive as would be applied to correct for blending. For comparison, we fit light curves from the PDC photometry and find our results do not change significantly. 

Our larger $\rhocirc$ and smaller $\bar{b}$ cannot be accounted for by dilution, which would produce the opposite effect \citep{kipp10}. Nor is the difference a result of our different prior on $\rhocirc$ or different methods of combining the posteriors from multiple transits { (Section \ref{subsec:dur})}. The difference could be due to different treatments of correlated noise. We can use $\rhocirc$ as a reality check for our derived values. \citet{nesv12} note that the TTVs constrain Kepler-46b's eccentricity to be very small and therefore $\rhocirc$ should match $\rho_\star$. We compute an updated value $\rho_\star$ by fitting the Dartmouth isochrones \citep{dott08} to \citet{nesv12}'s spectroscopic parameters and the Gaia parallax and magnitude \citep{gaia16,gaia18}, following \citet{daws19}. We find $R_\star = 0.833^{+0.020}{-0.013} R_\odot$, $M_\star = 0.89^{+0.02}_{-0.03} M_\odot$, and $\rho_\star = 1.54^{+0.10}{-0.16} \rho_\odot$, in good agreement with our light curve stellar density. 

Ultimately the small but significant differences in our parameters from those of \citet{nesv12} do not affect the main conclusion -- that the impact parameter is changing modestly -- except possibly to raise the concern that change we detect might be caused by dilution or correlated noise. The fact that our impact parameter is declining steadily over four years rather than oscillating from quarter to quarter gives us some confidence that the change is astrophysical. 

\begin{figure}
\begin{center}
\includegraphics[width=\columnwidth]{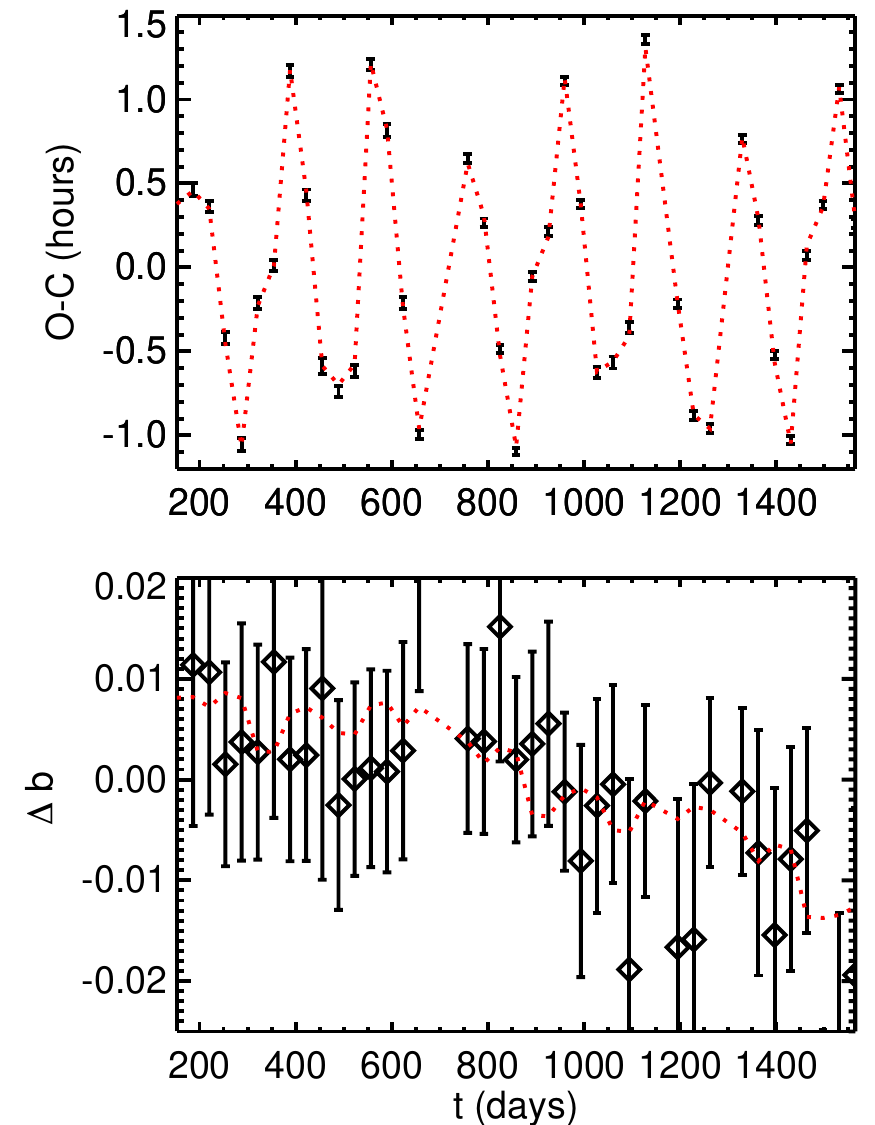}
\caption{Kepler-46b. Top: Observed minus calculated mid-transit times (black) and model (red dotted). Bottom: Change in impact parameter. 
\label{fig:model872}
}
\end{center}
\end{figure}

Figure \ref{fig:model872} shows an example of a dynamical model that provides a good fit { ($\chi^2 = 51$ for 66 degrees of freedom)} to the mid transit times, average impact parameter, and change in impact parameter. We use the stellar parameters derived above; the other astrocentric model parameters at epoch 55053.2826 BJD are $M_b =1.0 M_{\rm Jup}$, $P_b = 33.568$ days, $e_b=0.022$, $\omega_b=0$, $\Omega_b=0$, $i_b=89.26^\circ$, and mean anomaly $M_b=89.51^\circ$ and  $M_c =0.36 M_{\rm Jup}$, $P_c = 57.402$ days, $e_c=0.037$, $\omega_c=11^\circ$, $\Omega_b=-0.10^\circ$, $i_c=90.23^\circ$, and mean anomaly $M_c=353.7^\circ$ in the transit coordinate system with sky in the X-Y plane and +Z axis pointing at the observer (e.g., \citealt{winn10}). The mutual inclination $0.97^\circ$ is consistent with \citet{saad17}'s $0.43^\circ~^{+0.40}_{-0.46} $ to within two sigma. The transit duration variations computed from the model are dominated by changes in impact parameter. Future dynamical { modeling} can more thoroughly explore to what extent the detection of this change in impact parameter allows for better constraints on planet parameters, including the mutual inclination. { We recommend that a full exploration of parameter space using the dynamical model fit the durations rather than impact parameters to avoid applying the same prior twice (as discussed in Section \ref{subsec:dur}).} We also recommend full joint dynamical-photometry modeling\footnote{We avoid the common term ``photodynamical'' model because the term has sometimes refereed to a joint dynamical-photometry model (e.g., \citealt{mill17}) and sometimes to a two step (first photometry, then dynamical) model (e.g., \citealt{nesv14}).} for this system.

\subsection{Kepler-108b and c, a mutually inclined planetary system}
\label{subsec:108}
The Kepler-108 system contains two transiting warm Saturns on orbits mutually inclined by $I (^\circ) = 24^{+11}_{-8} $ \citep{mill17}. Both transiting planets exhibit TTVs. Moreover, planet c exhibits clear TDVs, with the transit duration changing by almost an hour over the course of about three years. \citet{mill17} note that planet b may also have TDVs but the change in duration is smaller and less significant (their Fig. 1). \citet{mill17} fit the light curves using a joint dynamical-photometry model: an $N$-body integrator models the orbits of the planets and star, and each light curve model is generated based on the planet's instantaneous orbit. This approach naturally generates TDVs in the case of non-coplanar planets. More { often,} studies first fit the light curves using light curve model parameters and subsequently fit a dynamical model to these light curve parameters  (e.g., \citealt{daws14a}). The latter two step approach is faster but the results can be sensitive to the choice of light curve parameters and their priors (e.g., as we have demonstrated here).

\begin{figure}
\begin{center}
\includegraphics[width=\columnwidth]{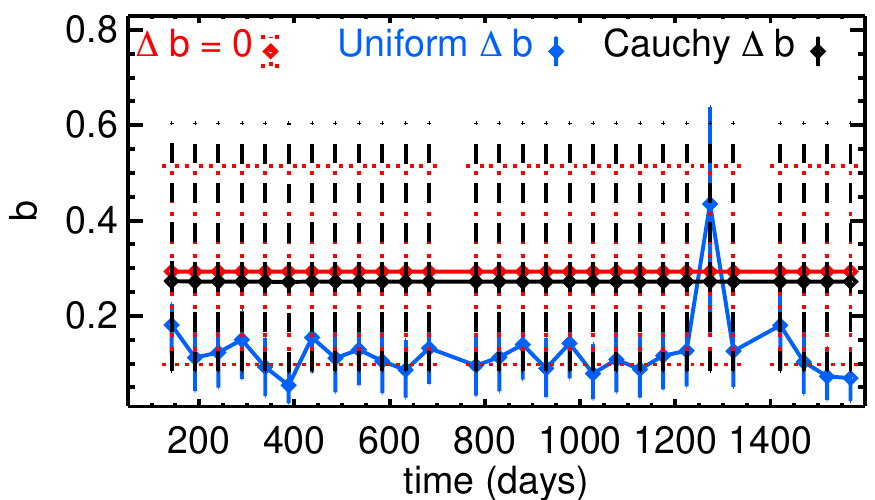}
\includegraphics[width=\columnwidth]{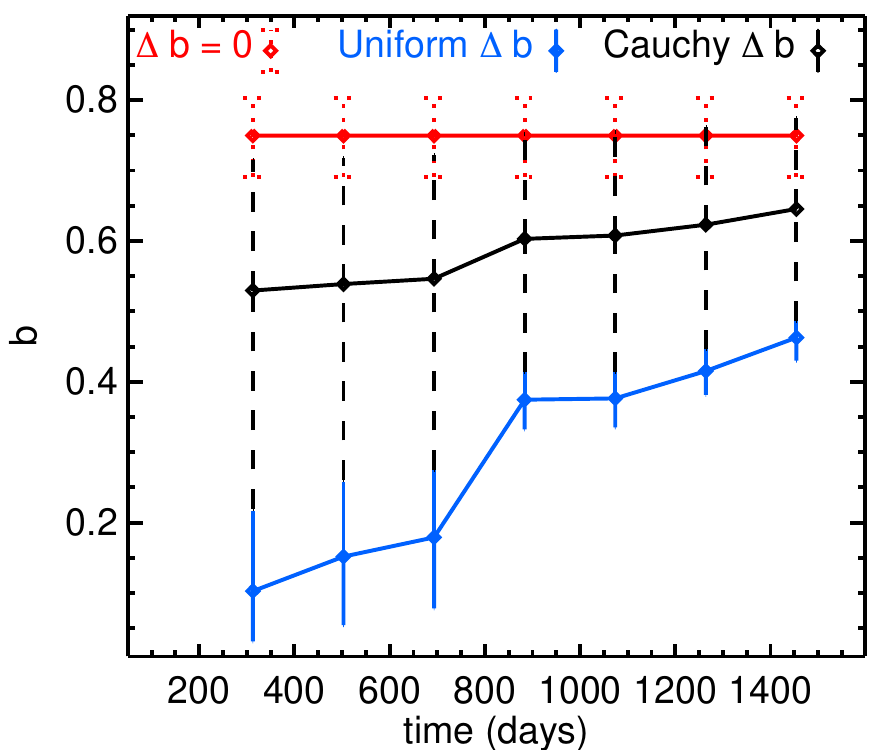}
\includegraphics[width=\columnwidth]{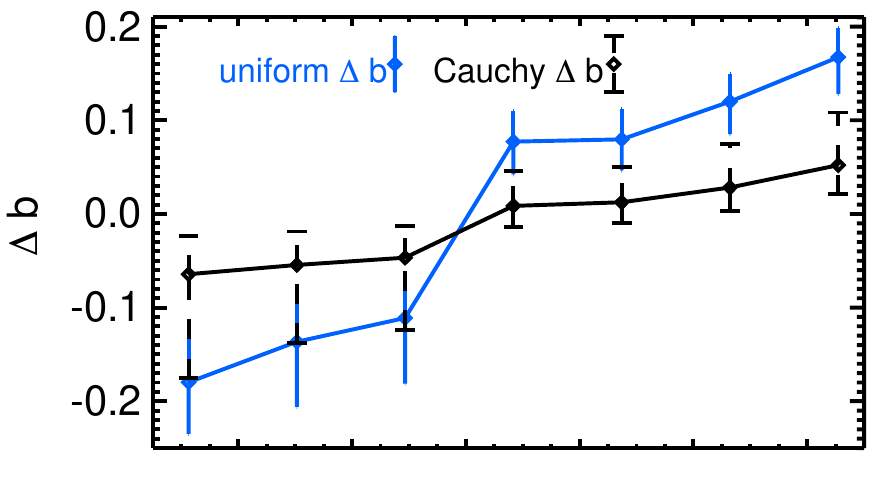}
\caption{Impact parameter vs. time fit for the Kepler-108 dataset (flux vs. time) for Kepler-108b (left and Kepler-108c (right) forcing the impact parameter to be constant from transit to transit (red), allowing the impact parameter to vary with a uniform prior on the change (blue), and allowing the impact parameter to vary with a more appropriate (Section \ref{sec:miti}) Cauchy prior on the change (black). With our favored Cauchy prior (black), we do not detect a change in impact parameter for Kepler-108b. We do detect a change in impact parameter for Kepler-108c but more modest than would be inferred with a uniform prior (blue). D20 depicts the credible interval for the black posterior and MF17 the value reported by \citet{mill17}.
\label{fig:108b}
}
 \end{center}
\end{figure}

\begin{figure*}
\begin{center}
\includegraphics[width=\columnwidth]{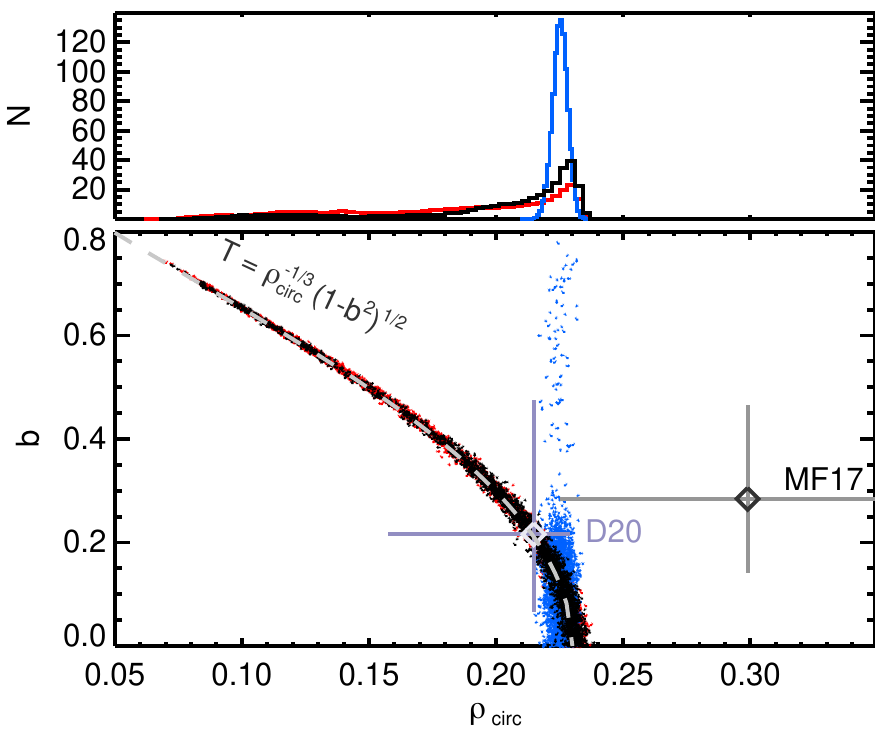}
\includegraphics[width=\columnwidth]{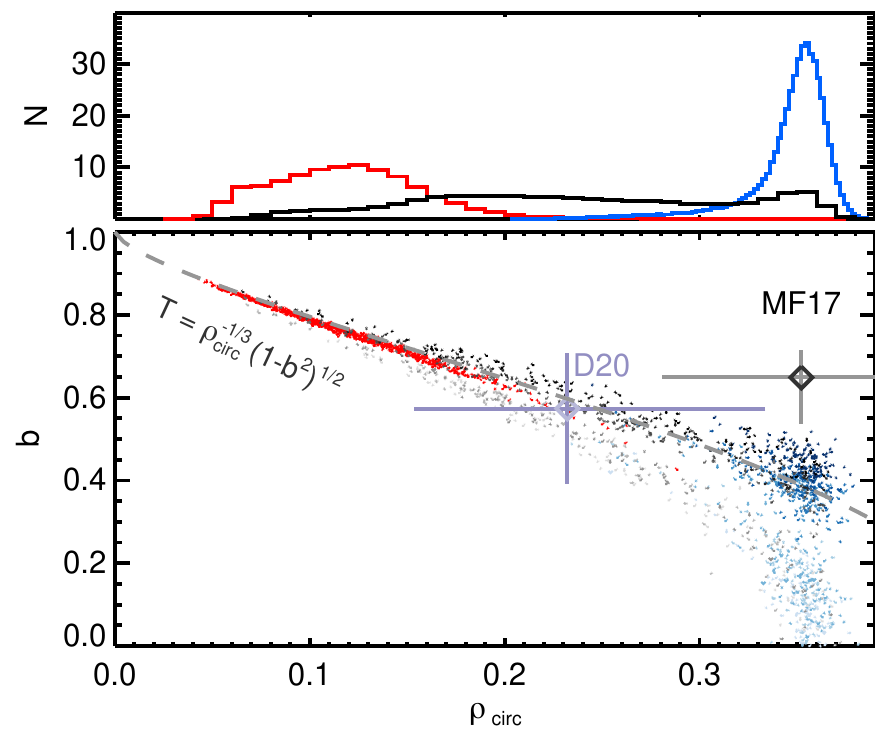}
\caption{Top: marginal posterior distribution for $\rhocirc$ when a Cauchy (red) or uniform (blue) prior is imposed on the change in impact parameter for Kepler-108b (left) and 108c (right). Bottom: Two dimensional posterior distribution for $b$ vs. $\rhocirc$.
\label{fig:108t}
}
 \end{center}
\end{figure*}

In Fig. \ref{fig:108b} and \ref{fig:108t}, we plot the results of our light curve fits for Kepler-108. Following \citet{mill17}, we account for dilution from a background star by including an extra parameter, the dilution factor. { We set a uniform prior on the dilution factor. We fit the light curves of both planets simultaneously, with shared values for the stellar limb darkening parameters, noise parameters, and dilution factor.} We report our best-fit parameters in Table \ref{tab:108}. Consistent with \citet{mill17}, we detect a change in impact parameter of Kepler-108c. The change scale is $\gamma=0.04^{+0.05}_{-0.02}$. If we had allowed the impact parameter to vary uniformly and independently, we would have overestimated the magnitude of the change (Fig. \ref{fig:108b}).  We do not detect a significant change in impact parameter of Kepler-108b. 

{ We note as a caveat that when using the alternative approach of fitting individual parameters to each transit (Section \ref{subsec:dur}), if we fit a common dilution factor, we deduce very little dilution, inconsistent with our other fit and \citet{mill17}. This result underscores our recommendation that the such fits (i.e., with individual parameters for each transit) should only be used to obtain transit times and durations to feed into dynamical models, not to infer other parameters.}

\begin{deluxetable}{rrl}
\tabletypesize{\footnotesize}
\tablecaption{Planet Parameters for { Kepler}-108b and c Derived from the Light-curves \label{tab:108}}
\tablewidth{0pt}
\tablehead{
\colhead{Parameter}    & \colhead{Value\tablenotemark{a}}}
\startdata
\hline
\\
Kepler-108b\\
Planet-to-star radius ratio, $R_{p}/R_{\star}$   				&0.067 &$^{+0.008}_{-0.017}$ \\
Light curve stellar density, $\rhocirc$  [$\rho_\odot$]  					&0.215 &$^{+0.134}_{-0.075}$ \\
Average impact parameter, $\bar{b}$ 					&0.21 &$^{+0.32}_{-0.15}$ \\
{ Impact parameter change scale,} $\gamma$  ($10^{-5}$)&3&$^{+850}_{-3}$ \\
\hline\\
Kepler-108c \\
Planet-to-star radius ratio, $R_{p}/R_{\star}$   				&0.057 &$^{+0.007}_{-0.015}$ \\
Light curves stellar density, $\rhocirc$  [$\rho_\odot$]  					&0.24 &$^{+0.10}_{-0.09}$ \\
Average impact parameter, $\bar{b}$ 					&0.57 &$^{+0.14}_{-0.18}$ \\
{ Impact parameter change scale,} $\gamma$&0.04&$^{+0.05}_{-0.02}$ \\
\hline\\
Limb darkening coefficient, $q_{1}$ 						& $0.30$& $\pm 0.07$ 		\\
Limb darkening coefficient, $q_{2}$ 						& $0.37$&$^{+0.10}_{-0.09}$	 		\\
Dilution factor & 0.70 &$^{+0.06}_{-0.27}$	 		\\
Red noise, short-cadence, $\sigma_r$	[ppm]		& 2330 & $\pm150$  		\\
White noise, short-cadence $\sigma_w$	[ppm]		& 460.&$\pm 2$  		\\
Red noise, long-cadence $\sigma_r$		[ppm]	& 360&$\pm 20$  		\\
White noise, long-cadence $\sigma_w$	[ppm]		& 88.7&$\pm 1.6$  													\enddata
\tablenotetext{a}{The uncertainties represent the 68.3\% credible interval about the median of the posterior distribution.}
\end{deluxetable}

Our inferred parameters in Table \ref{tab:108} are consistent with those of \citet{mill17}'s mutually inclined fit.   \citet{mill17} found an average impact parameter of $b=0.28^{+0.18}_{-0.14}$ for Kepler-108b and $b=0.65^{+0.06}_{-0.11}$ for Kepler-108c (Sean Mills, personal communication, March 10th 2017). Their { average} scaled planet-star separation corresponds to { $\rhocirc=0.30^{+0.30}_{-0.07} \rho_\odot$} for Kepler-108b and { $\rhocirc=0.35^{+0.14}_{-0.07} \rho_\odot$}  (Sean Mills, personal communication, March 10th 2017). Generally our uncertainties are larger. Our larger uncertainties may arise because we include noise parameters, including correlated noise, in our inference. Another possibility is that \citet{mill17} obtain more precise values because the joint dynamical-photometry model naturally imposes constraints on the light curve parameters { (i.e., due to the limited possible variations in transit impact parameter allowed by the physical model)}.

Using a joint dynamical-photometry model like \citet{mill17} naturally imposes an appropriate prior on the change in impact parameter; { the transit speed can vary as well according to the dynamical model. Therefore this approach is not subject to bias described in Section 2.} We recommend the joint dynamical-photometry approach if computationally feasible. However, when it is not computationally feasible due to a large dataset, the need to account for correlated noise, or a large sample size of planets, we recommend the approach presented here using the Cauchy prior on change in impact parameter.

\subsection{KOI-319.01, a transiting warm Jupiter perturbed by a non-transiting warm Saturn or warm Neptune}
\label{subsec:319}
\citet{nesv14} found that KOI-319.01 exhibits large TTVs caused by a non-transiting warm Saturn or warm Neptune. They detected fluctuating TDVs that are not consistent with the dynamical model, which predicts a constant or linearly drifting TDVs. \citet{nesv14} proposed that their TDV errors may be underestimated or may be caused by an unmodeled effect. { To assess the TDVs, \citet{nesv14} fit the data using a model in which each transit had its own $\rhocirc$, $b$, and $R_p/R_\star$  (Section \ref{subsec:dur}). Their analysis  was not subject to the bias described in Section \ref{sec:orig}.} 

\begin{figure}
\begin{center}
\includegraphics[width=\columnwidth]{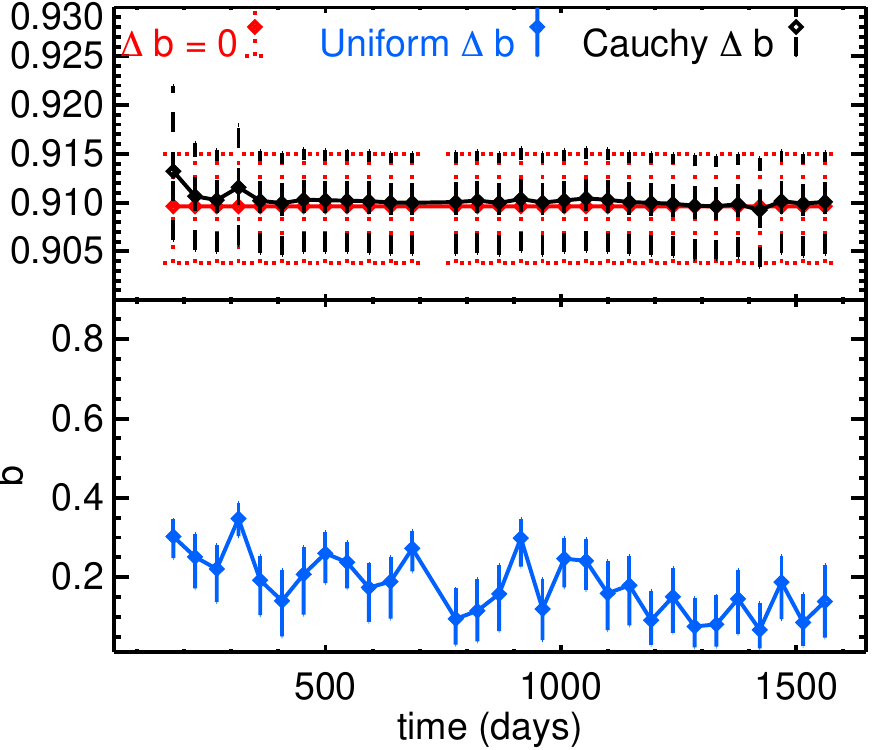}
\caption{Impact parameter vs. time fit from the KOI-319 dataset (flux vs. time) forcing the impact parameter to be constant from transit to transit (red), allowing the impact parameter to vary with a uniform prior on the change (blue), and allowing the impact parameter to vary with a more appropriate (Section \ref{sec:miti}) Cauchy prior on the change (black). 
\label{fig:319b}
}
\end{center}
\end{figure}

\begin{figure}
\begin{center}
\includegraphics[width=\columnwidth]{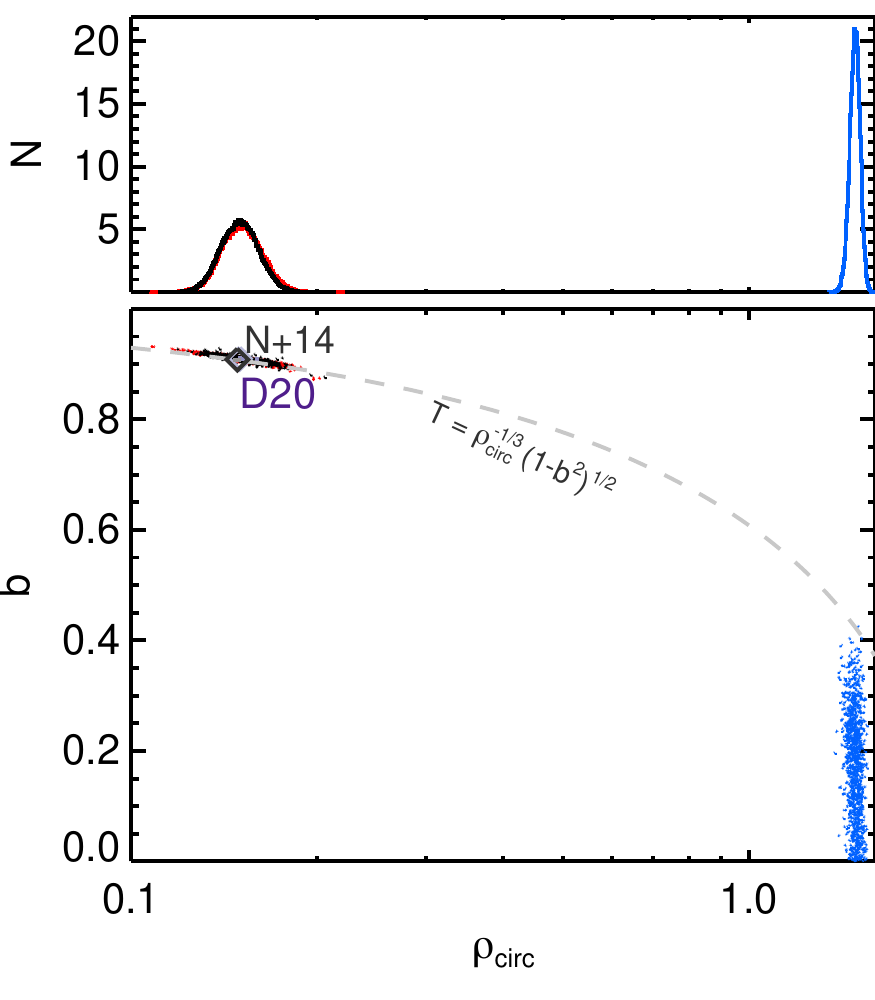}
\caption{Top: marginal posterior distribution for $\rhocirc$ when a Cauchy (red) or uniform (blue) prior is imposed on the change in impact parameter for the KOI-319 dataset. Bottom: Two dimensional posterior distribution for $b$ vs. $\rhocirc$. Best fit values from \citet{nesv14} are indicated. D20 depicts the credible interval for the black posterior.
\label{fig:319t}
}
\end{center}
\end{figure}

\begin{figure}
\begin{center}
\includegraphics[width=\columnwidth]{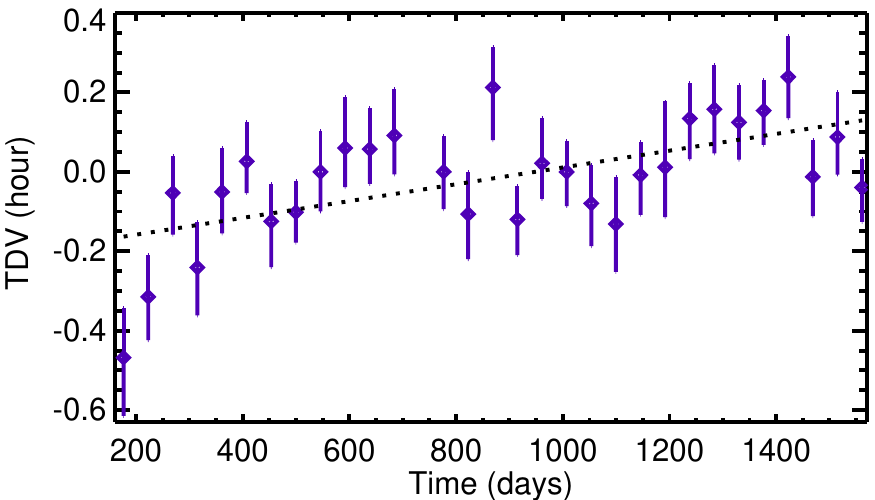}
\caption{ Change in transit duration for KOI-319, fitting each transit individually. We find that KOI-319b exhibits a drift in transit duration over the Kepler Mission, { but from the data alone we cannot definitively attribute this drift to a change in impact parameter.}
\label{fig:319dur}
}
\end{center}
\end{figure}

Our fit results are shown in Fig. \ref{fig:319b} and \ref{fig:319t} and Table \ref{tab:319}. Our parameters are very similar to and consistent with \citet{nesv14} except that our uncertainties are several times larger. Our fit without the appropriate prior shows some possible variation, but with an appropriate prior, the change is consistent with zero (Table \ref{tab:319}).   When we fit each transit individually following Section \ref{subsec:dur}, we see a drift in transit duration; because our error bars are larger, the changes are consistent with a linear drift (Fig. \ref{fig:319dur}). We conclude that the current data do not contain { sufficient evidence to definitively attribute the change in duration to a change in impact parameter}. We recommend { additional dynamical modeling of the duration variations and} full joint dynamical-photometry modeling of this system to tease out if and how the duration changes constrain the orbital parameters.

\begin{deluxetable}{rrl}
\tabletypesize{\footnotesize}
\tablecaption{Planet Parameters for KOI-319.01 Derived from the Light-curves \label{tab:319}}
\tablewidth{0pt}
\tablehead{
\colhead{Parameter}    & \colhead{Value\tablenotemark{a}}}
\startdata
\hline
\\
Planet-to-star radius ratio, $R_{p}/R_{\star}$   				&0.0471 &$^{+0.0020}_{-0.0013}$ \\
Light curves stellar density, $\rhocirc$  [$\rho_\odot$]  					&0.150 &$^{+0.011}_{-0.010}$ \\
Average impact parameter, $\bar{b}$ 					&0.910 &$^{+0.005}_{-0.005}$ \\
{ Impact parameter change scale,} $\gamma(10^{-5})$&5&$^{+54}_{-5}$ \\
Limb darkening coefficient, $q_{1}$ 						& $0.37$& $^{+0.06}_{-0.07}$ 		\\
Limb darkening coefficient, $q_{2}$ 						& $0.46$&$^{+0.36}_{-0.33}$	 		\\
Red noise, short-cadence, $\sigma_r$	[ppm]		& 1940 & $\pm140$  		\\
White noise, short-cadence $\sigma_w$	[ppm]		& 364&$\pm 2$  		\\
Red noise, long-cadence $\sigma_r$		[ppm]	& 370&$\pm20$  		\\
White noise, long-cadence $\sigma_w$	[ppm]		& 88&$\pm 2$  													\enddata
\tablenotetext{a}{The uncertainties represent the 68.3\% credible interval about the median of the posterior distribution.}
\end{deluxetable}

\subsection{Kepler-88b, a warm Neptune perturbed by a non-transiting, nearly coplanar warm Jupiter}
\label{subsec:142}
Kepler-88b is a warm Neptune perturbed by a non-transiting, nearly coplanar warm Jupiter in a 2:1 orbital resonance \citep{nesv13}. Kepler-88b is not the type of planet our approach is designed for: rather than being a warm Jupiter with a well-separated companion that causes nodal precession, Kepler-88b is a Neptune with a nearby massive resonant companion that can cause significant changes to the longitude of periapse (and hence $\rhocirc$) on a short timescale. As such it makes an interesting test case for our approach, which assumes that only the impact parameter can change detectably.

\citet{nesv13} found small but significant TDVs for Kepler-88b, the first TDVs { due to planet-planet interactions} detected to our knowledge. { To assess the TDVs, \citet{nesv13} fit the data using a model in which each transit had its own $\rhocirc$, $b$, and $R_p/R_\star$  (Section \ref{subsec:dur}). Their analysis  was not subject to the bias described in Section \ref{sec:orig}.}  The companion is well-characterized from the TTVs alone and a dynamical fit to only the TTVs predicts the TDVs too. The TDVs are primarily caused by changes in the transit speed (i.e., $\rhocirc$), rather than the impact parameter. \citet{weis19} recently followed up the system with the radial velocity method and performed { joint} dynamical-photometry modeling on the combined dataset; they also found significant TDVs.

\begin{figure}
\begin{center}
\includegraphics[width=\columnwidth]{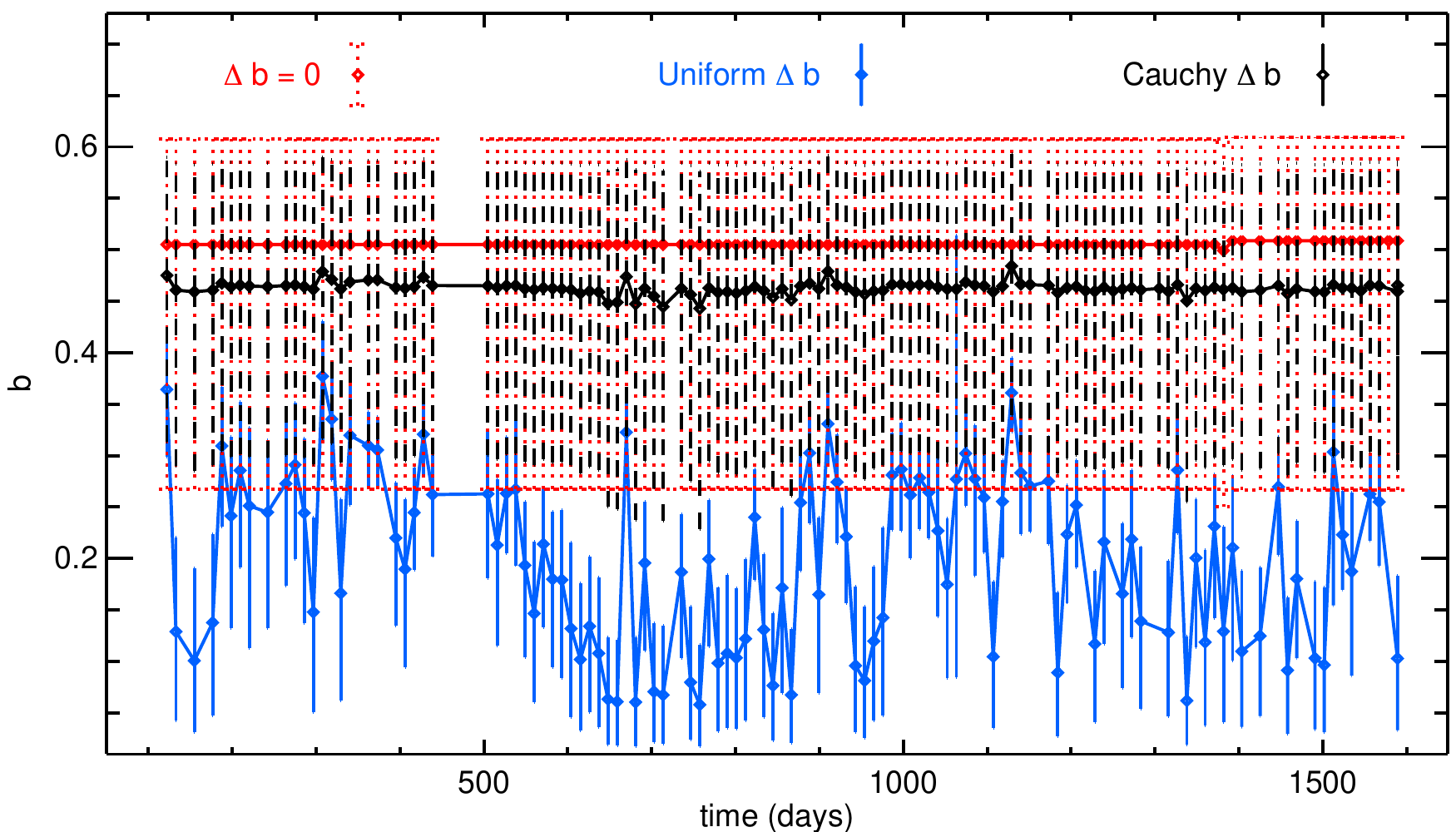}
\caption{Impact parameter vs. time fit from the Kepler-88 dataset (flux vs. time) forcing the impact parameter to be constant from transit to transit (red), allowing the impact parameter to vary with a uniform prior on the change (blue), and allowing the impact parameter to vary with a more appropriate (Section \ref{sec:miti}) Cauchy prior on the change (black). Using the Cauchy prior, we do not detect a significant change in impact parameter.
\label{fig:142b}
}
\end{center}
\end{figure}

\begin{figure}
\begin{center}
\includegraphics[width=\columnwidth]{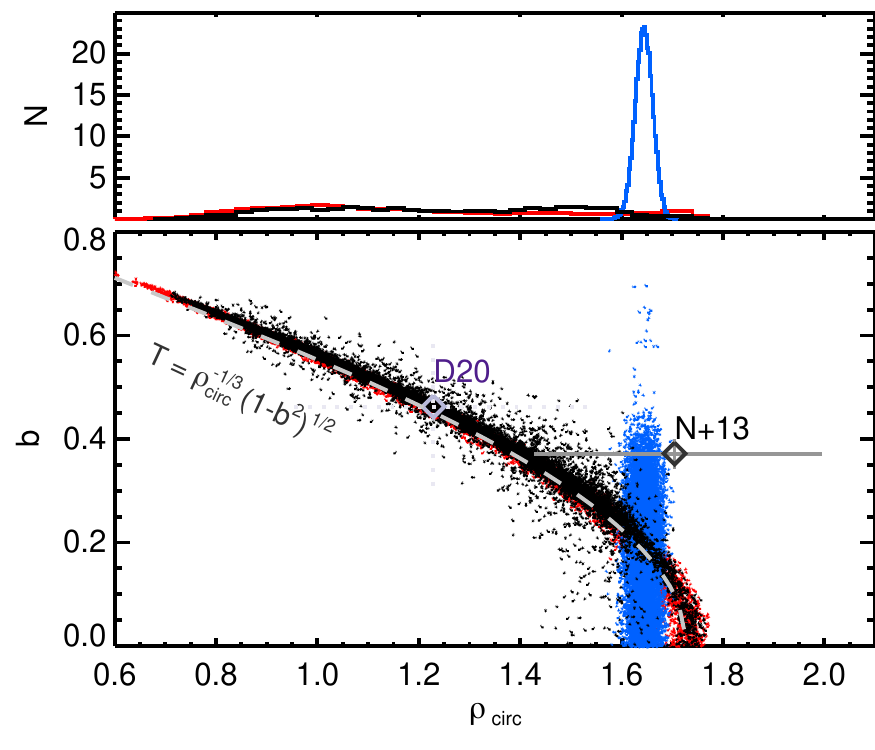}
\caption{ Top: marginal posterior distribution for $\rhocirc$ when a Cauchy (red) or uniform (blue) prior is imposed on the change in impact parameter for the Kepler-88 dataset. Bottom: Two dimensional posterior distribution for $b$ vs. $\rhocirc$. D20 depicts the credible interval for the black posterior and N+13 the value reported by \citet{nesv13}.
\label{fig:142t}
}
\end{center}
\end{figure}

The results from our fits are shown in Fig. \ref{fig:142b} and \ref{fig:142t} and tabulated in Table \ref{tab:88}. Without an appropriate prior for the change in impact parameter (i.e., blue), we might erroneously conclude that the impact parameter is changing. An appropriate prior (black) allows us to correctly deduce that the impact parameter does not { change} detectably over the timespan of the dataset. As \citet{nesv13} and \citet{weis19} simulate, the impact parameter can change over a much longer timescale such that eventually Kepler-88b no longer transits, but the Kepler dataset is not long and/or precise enough to detect a change. We believe that the mutual inclination measurement is primarily coming from the TTVs rather than the TDVs, though the TDVs may be contributing an upper limit. Our parameters in Table \ref{tab:88} are consistent with \citet{nesv13} and \citet{weis19} to within the uncertainties.

\begin{deluxetable}{rrl}
\tabletypesize{\footnotesize}
\tablecaption{Planet Parameters for Kepler-88b Derived from the Light-curves \label{tab:88}}
\tablewidth{0pt}
\tablehead{
\colhead{Parameter}    & \colhead{Value\tablenotemark{a}}}
\startdata
\hline
\\
Planet-to-star radius ratio, $R_{p}/R_{\star}$   				&0.0353 &$^{+0.0008}_{-0.0006}$ \\
Light curves stellar density, $\rhocirc$  [$\rho_\odot$]  					&1.2 &$^{+0.3}_{-0.3}$ \\
Average impact parameter, $\bar{b}$ 					&0.46 &$^{+0.12}_{-0.17}$ \\
{ Impact parameter change scale,} $\gamma(10^{-4})$&7&$^{+83}_{-7}$ \\
Limb darkening coefficient, $q_{1}$ 						& $0.47$& $^{+0.07}_{-0.06}$ 		\\
Limb darkening coefficient, $q_{2}$ 						& $0.30$&$^{+0.08}_{-0.06}$	 		\\
Red noise, short-cadence, $\sigma_r$	[ppm]		& 870 & $\pm90$  		\\
White noise, short-cadence $\sigma_w$	[ppm]		& 551.5&$\pm 1.0$  		\\
Red noise, long-cadence $\sigma_r$		[ppm]	& 80&$\pm50$  		\\
White noise, long-cadence $\sigma_w$	[ppm]		& 108&$\pm 3$  													\enddata
\tablenotetext{a}{The uncertainties represent the 68.3\% credible interval about the median of the posterior distribution.}
\end{deluxetable}

\subsection{TOI-216{ b and c, a pair of warm Jupiters}}
\label{subsec:216}
TOI-216 hosts a pair of transiting warm, large exoplanets in or near the 2:1 orbital resonance \citep{kipp19,daws19}. The inner planet's grazing transit configuration makes its transit durations particularly sensitive to a small precession of the longitude of ascending node. Moreover, based on the planets' impact parameters, \citet{daws19} found a minimum mutual inclination of $1.8^{+0.2}_{-0.2}$ degrees. { Neither previous study investigated changes in impact parameter or transit duration variations. We fit the TESS simple aperture photometry from MAST together with the ground-based light curves presented in \citet{daws19}. We fit the light curves of both planets simultaneously, with shared values for the stellar limb darkening parameters and noise parameters.} We do not detect a significant change in impact parameter for either planet (Table \ref{tab:216}). We recommend continued observations from the ground to monitor the inner planet for changes in impact parameter.

\begin{deluxetable}{rrl}
\tabletypesize{\footnotesize}
\tablecaption{Planet Parameters for TOI-216{ b and c} from the Light-curves \label{tab:216}}
\tablewidth{0pt}
\tablehead{
\colhead{Parameter}    & \colhead{Value\tablenotemark{a}}}
\startdata
\hline
\\
TOI-216b\\
Planet-to-star radius ratio, $R_{p}/R_{\star}$   				&0.11 &$^{+0.04}_{-0.03}$ \\
Light curves stellar density, $\rhocirc$  [$\rho_\odot$]  					&1.1 &$^{+0.3}_{-0.2}$ \\
Average impact parameter, $\bar{b}$ 					&1.01 &$^{+0.05}_{-0.05}$ \\
{ Impact parameter change scale,} $\gamma(10^{-5})$&1&$^{+4}_{-1}$ \\
Planet-to-star radius ratio, $R_{p}/R_{\star}$   				&0.1230 &$^{+0.0007}_{-0.0007}$ \\
Light curves stellar density, $\rhocirc$  [$\rho_\odot$]  					&1.73 &$^{+0.04}_{-0.05}$ \\
Average impact parameter, $\bar{b}$ 					&0.13 &$^{+0.07}_{-0.07}$ \\
{ Impact parameter change scale,} $\gamma(10^{-8})$&3&$^{+1969}_{-3}$ \\
TOI-216c\\
Planet-to-star radius ratio, $R_{p}/R_{\star}$   				&0.1230 &$^{+0.0007}_{-0.0007}$ \\
Light curves stellar density, $\rhocirc$  [$\rho_\odot$]  					&1.73 &$^{+0.04}_{-0.05}$ \\
Average impact parameter, $\bar{b}$ 					&0.14 &$\pm0.07$ \\
{ Impact parameter change scale,} $\gamma(10^{-8})$&3&$^{1969}_{-3}$ \\
System\\
TESS limb darkening coefficient, $q_{1}$ 						& $0.32$& $^{+0.10}_{-0.08}$ 		\\
TESS limb darkening coefficient, $q_{2}$ 						& $0.47$&$^{+0.15}_{-0.11}$	 		\\
TESS red noise $\sigma_r$	[ppm]		& 3700 & $^{+700}_{-600}$  		\\
TESS white noise $\sigma_w$	[ppm]		& 2481&$\pm 14$  		\\													\enddata
\tablenotetext{a}{The uncertainties represent the 68.3\% credible interval about the median of the posterior distribution.}
\end{deluxetable}

\section{Summary}
\label{sec:disc}

Changes in a transiting planet's impact parameter can constrain the mutual inclinations of planetary systems, including mutual inclinations between the transiting planets and non-transiting companions. Evidence for changes in impact parameters can be evaluated in existing Kepler and TESS data, future TESS data, and planned PLATO data. We presented a demonstration of a problem of incorrect inference of changes in impact parameter from transit light curves (Section \ref{sec:orig}) and two approaches for mitigating the problem (Section \ref{sec:miti}).

We applied our results to systems from the literature (Sections \ref{sec:appl1} and \ref{sec:appl2}), { most of which were not subject to the bias described here in their previous studies.} We discovered evidence for a change in impact parameter for Kepler-46b (Section \ref{subsec:872}). We confirmed changes in impact parameter for two planets with detected transit duration variations (TDVs), Kepler-639b (Section \ref{subsec:693}) and Kepler-108b (Section \ref{subsec:108}). We confirmed no evidence for a change in impact parameter for Kepler-448b (Section \ref{subsec:448}) and TOI-216 b { and c} (Section \ref{subsec:216}); for the ambiguous cases of Kepler-419b (Section \ref{subsec:419}), Kepler-108c (Section \ref{subsec:108}), and KOI-319.01 (Section \ref{subsec:319}){, which exhibits transit duration variations that cannot be definitively attributed to a change in impact parameter from the data alone}; and for Kepler-88b (Section \ref{subsec:142}).

The ideal approach for fitting light curves is to simultaneously use a joint photometry-dynamics model and a regression approach that accounts for correlated noise, but in practice, there is a high computational cost to doing both simultaneously off the bat. We recommend the following { approaches} to ensure the results are robust to parameter choices and model assumptions without requiring unrealistic computation times:
\begin{enumerate}
\item { To identify changes in impact parameter and/or to obtain a robust $\rhocirc$ posterior in the presence of possible changes in impact parameter:} fit the light curves with individual transit times; individual impact parameters for each transit; and a Cauchy prior on $\gamma$, the scale of the change in impact parameter (Section \ref{subsec:prior}). Specifically, we recommend fitting { mid transit times $t_{i}$} and { changes in impact parameter} $\Delta b_i = b_i - \bar{b}$ for each of $i$ transits and a joint $R_p/R_\star$, $\rhocirc$, average impact parameter\footnote{Fitting $\bar{b}$ as an extra parameter allows us to easily obtain the posterior for this quantity, and $\Delta b_i$ is often more precisely constrained than $b_i$.} $\bar{b}$, impact parameter change scale $\gamma$, and limb darkening and noise parameters among all transits. Include Eqn. \ref{eqn:cauchy} in the { prior}. Use an approach that accounts for correlated noise and does not require pre-detrending, such as a wavelet likelihood combined with linear trends fit to each light curve segment or Gaussian process regression. Use the posteriors for the noise parameters to identify if: a) white noise dominates, b) only long timescale correlated noise (i.e., a linear trend or polynomial) is important, or c) short timescale noise is important { too} and therefore a wavelet or Gaussian process likelihood (or an alternative approach) should be included.
\item { If the goal is to obtain transit durations for use in a dynamical model, fit individual $t_{i}$, $R_{p,i}/R_\star$, ${\rhocirc}_i$, and $b_i$ (and joint values only for limb darkening and noise parameters), applying the prior in Eqn. \ref{eqn:like} to preserve a uniform prior on transit durations (Section \ref{subsec:dur}). Compute the transit durations $T_i$ from Eqn. \ref{eqn:time} (modifying in the case of grazing transits). Do not use this approach to obtain posteriors for $R_p/R_\star$, $\rhocirc$, and $b$; posteriors for these values should be obtained using the first approach or, less precisely, fitting parameters jointly to all light curves or a binned, phase-folded light curve. See Section \ref{subsec:dur} for further discussion.}
\item { If fitting a dynamical model, use the transit times, average impact parameter $\bar{b}$, $\rhocirc$, and changes in impact parameter $\Delta b_i$ from step 1 to identify a dynamical model as a starting point (e.g., as we perform for Kepler-46b in Section \ref{subsec:872}). Then directly fit transit times $t_i$ and durations $T_i$ from Step 2 to explore the parameter space for the dynamical model. As discussed in \ref{subsec:dur}, it is important to fit the transit durations instead of changes in impact parameter to avoid applying priors on the impact parameter twice.}
\item { If computationally feasible,} fit a full joint photometry-dynamics model to the light curves and compare to the previous step to check for consistency. Use the results of Step 1 to assess if and how correlated noise should be accounted for. If short timescale correlated noise needs to be accounted for yet it is not computationally feasible to do so, compare $R_p/R_\star$ { from Step 1} to get a sense for how much the uncertainties may be underestimated.
\end{enumerate}

Ultimately the presence or absence of detectable changes in impact parameter can help constrain the origins of warm Jupiters. { More consideration is needed on the best way to incorporate grazing transits into population studies: they can be quite sensitive to small changes in impact parameter but often have poorly constrained radii. For example, \citet{daws15} included them in their population weighted by their probability of having a Jupiter-like radius, but such an approach is sensitive to the assumed prior on radius.} It is important not to exclude nearly grazing transits, as they are particularly sensitive to small changes in impact parameter. Ultimately, since changes in impact parameter manifest as long timescale drift, Plato can play an essential role by following up the Kepler field and revisiting other fields over a long observational baseline. TESS warm Jupiters can be followed up from the ground (e.g., \citealt{daws19}) or by CHEOPs to increase the observational baseline.  In combination with ground-based follow up, we can also investigate whether orbital architectures correlate with stellar metallicity or other properties.

\section*{Acknowledgments}
{ I thank the referee,  Kento Matsuda, for a particularly helpful, thoughtful report that greatly improved the paper.} I thank Daniel Fabrycky, Eric Ford, Tom Loredo, Sean Mills, Darin Ragozzine, Leslie Rogers, and Angie Wolfgang for helpful discussions. I gratefully acknowledge the 2016--2017 Program on Statistical, Mathematical and Computational Methods for Astronomy Astrophysical Populations working group and the 2013 Modern Statistical and Computational Methods for Analysis of Kepler Data Bayesian Characterization of Exoplanet Populations Working Group and Noise and Detrending (No More Tears) Working Group.

I gratefully acknowledge support from grant NNX16AB50G awarded by the NASA Exoplanets Research Program and the Alfred P. Sloan Foundation's Sloan Research Fellowship. The Center for Exoplanets and Habitable Worlds is supported by the Pennsylvania State University, the Eberly College of Science, and the Pennsylvania Space Grant Consortium.  This material was based upon work partially supported by the National Science Foundation under Grant DMS-1127914 to the Statistical and Applied Mathematical Sciences Institute and under Grant No. NSF PHY-1748958. Any opinions, findings, and conclusions or recommendations expressed in this material are those of the author(s) and do not necessarily reflect the views of the National Science Foundation. This research or portions of this research were conducted with Advanced CyberInfrastructure computational resources provided by The Institute for Computational and Data Sciences at The Pennsylvania State University (https://ics.psu.edu).

I include data collected by the \kep mission, funded by the NASA Science Mission directorate, and thank the \kep team for producing these data sets. Light curves were downloaded from the Mikulski Archive for Space Telescopes (MAST). Some data were obtained from the NASA Exoplanet Archive, operated by Caltech, under contract with the NASA Exoplanet Exploration Program.   This work has made use of data from the European Space Agency (ESA) mission {\it Gaia} (\url{https://www.cosmos.esa.int/gaia}), processed by the {\it Gaia} Data Processing and Analysis Consortium (DPAC, \url{https://www.cosmos.esa.int/web/gaia/dpac/consortium}). Funding for the DPAC has been provided by national institutions, in particular the institutions participating in the {\it Gaia} Multilateral Agreement. Resources supporting this work were provided by the NASA High-End Computing (HEC) Program through the NASA Advanced Supercomputing (NAS) Division at Ames Research Center for the production of the SPOC data products.

I thank the \TESS Mission team and follow up working group for the valuable dataset. We acknowledge the use of public \TESS Alert data from pipelines at the \TESS Science Office and at the \TESS Science Processing Operations Center.  This paper includes data collected by the \TESS mission, which are publicly available from the Mikulski Archive for Space Telescopes (MAST).  This research has made use of the Exoplanet Follow-up Observation Program website, which is operated by the California Institute of Technology, under contract with the National Aeronautics and Space Administration under the Exoplanet Exploration Program.

 \software{TAP\citep{gaza12}}
 
\bibliographystyle{apj}
\bibliography{ms}
\clearpage
\appendix

\begin{figure*}[b]
\begin{center}
\includegraphics[width=.5\columnwidth]{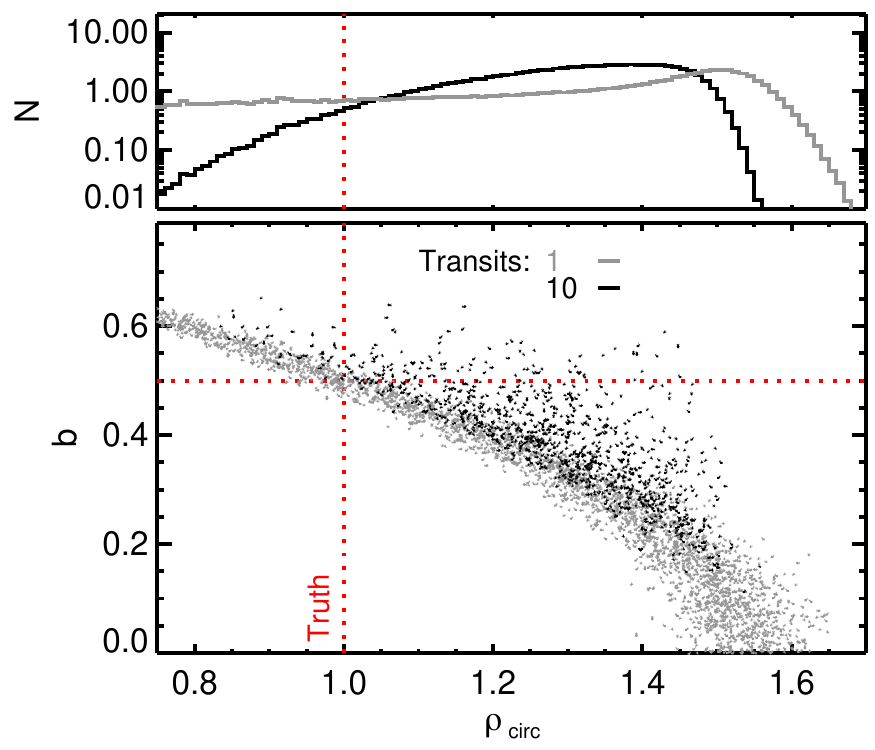}\includegraphics[width=.5\columnwidth]{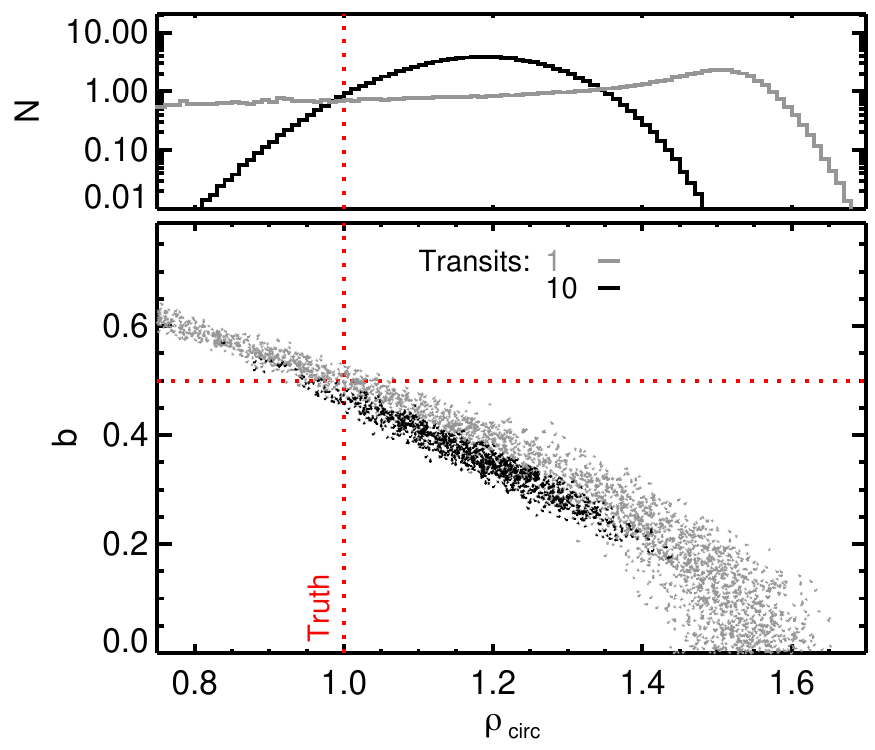}
\caption{Averaging the posteriors samples from individual transits using the median (left) or mean (right) shifts the posterior away from the truth as more transits are added. Gray corresponds to the inference from a single transit for the marginal $\rhocirc$ posterior (top) and joint $(\rhocirc,b)$ posterior (bottom). Black corresponds to ten transits. In the left panel, the ten transit posterior is shifted away from the truth toward larger $b$ and larger $\rhocirc$. In the left panel, the right transit posterior is shifted away from the truth toward smaller $b$ and larger $\rhocirc$. 
\label{fig:indiv}
}
\end{center}
\end{figure*}

Fig. \ref{fig:indiv} demonstrates the bias introduced by averaging posterior samples across individual  transits.

\label{lastpage}

\end{document}